\documentclass[aps,prb,10pt,twocolumn,superscriptaddress,floatfix]{revtex4-1}

\usepackage{graphicx,graphics}
\usepackage{dcolumn}
\usepackage{amsmath,amssymb,amsfonts}
\usepackage{latexsym,verbatim}
\usepackage{bm}
\usepackage{color}
\usepackage{ulem}
\usepackage[percent]{overpic}
\usepackage[breaklinks=true,colorlinks,citecolor=blue,linkcolor=blue,urlcolor=blue]{hyperref}

\DeclareMathAlphabet\mathbfcal{OMS}{cmsy}{b}{n}

\newcommand{\bB}{{\bm B}}

\newcommand{\bJ}{{\bm J}}

\newcommand{\br}{{\bm r}}
\newcommand{\bv}{{\bm v}}

\newcommand{\sign}{\mathrm{sgn}}
\begin{document}
\title{Non-local transport and the Hall viscosity of 2D hydrodynamic electron liquids}
\author{Francesco M.D. Pellegrino}
\email{francesco.pellegrino@sns.it}
\affiliation{NEST, Scuola Normale Superiore, I-56126 Pisa,~Italy}
\author{Iacopo Torre}
\affiliation{NEST, Scuola Normale Superiore, I-56126 Pisa,~Italy}
\affiliation{Istituto Italiano di Tecnologia, Graphene Labs, Via Morego 30, I-16163 Genova,~Italy}
\author{Marco Polini}
\affiliation{Istituto Italiano di Tecnologia, Graphene Labs, Via Morego 30, I-16163 Genova,~Italy}
\affiliation{School of Physics \& Astronomy, University of Manchester, Oxford Road, Manchester M13 9PL, United Kingdom}
\begin{abstract}
In a fluid subject to a magnetic field the viscous stress tensor has a dissipationless antisymmetric component controlled by the so-called Hall viscosity. We here propose an all-electrical scheme that allows a determination of the Hall viscosity of a two-dimensional electron liquid in a solid-state device. 
\end{abstract}

\maketitle

\section{Introduction}
\label{sec:introduction}

Electron systems roaming in a crystal where the mean free path for electron-electron collisions is the shortest length scale of the problem can be described by conservation laws for macroscopic collective variables~\cite{gurzhi_spu_1968,dyakonov_prl_1993,dyakonov_prb_1995,dyakonov_ieee_1996,conti_prb_1999,govorov_prl_2004,muller_prb_2008,fritz_prb_2008,muller_prl_2009,
bistritzer_prb_2009,andreev_prl_2011,mendoza_prl_2011,svintsov_jap_2012,mendoza_scirep_2013,tomadin_prb_2013,tomadin_prl_2014,lucas_njp_2015,
torre_prb_2015_I,narozhny_prb_2015,briskot_prb_2015,torre_prb_2015,levitov_naturephys_2016,pellegrino_prb_2016,lucas_prb_2016,Levchenko_prb_2017}. Such non-perturbative hydrodynamic description relies on the knowledge of a small number of kinetic coefficients~\cite{landaufluidmechanics}, i.e.~the bulk $\zeta$ and shear $\eta$ viscosities and the thermal conductivity $\kappa$. In the presence of time-reversal symmetry, these fully determine the response of the electron system to slowly-varying 
external fields. However, when time-reversal symmetry is broken (for example due to the presence of an external magnetic field), a dissipationless term, controlled by the so-called Hall viscosity~\cite{avron_prl_1995,tokatly_prb_2007,tokatly_jpcm_2009,read_prb_2009,read_prb_2011,haldane_prl_2011,hoyos_prl_2012,bradlyn_prb_2012,sherafati_prb_2016,alekseev_prl_2016,cortijo_2DM_2016,scaffidi_prl_2017} $\eta_{\rm H}$, appears in the viscous stress tensor~\cite{landaufluidmechanics} $\sigma^\prime_{ij}$. In two spatial dimensions one has
\begin{equation}\label{eq:stress_Cartesian_indices}
\sigma^\prime_{ij} = \sum_{k, \ell} \eta_{ij, k\ell} v_{k\ell}~,
\end{equation}
where $i,j,k$ and $\ell$ denote Cartesian indices, $v_{k\ell} \equiv (\partial_{k} v_{\ell} +\partial_{\ell} v_{k})/2$, and $\eta_{ij, k\ell}$ is a rank-$4$ tensor, usually called ``viscosity'' tensor~\cite{bradlyn_prb_2012},
\begin{equation}\label{eq:viscosity_tensor}
\begin{split}
\eta_{ij, k\ell}  &\equiv \zeta \delta_{ij}\delta_{k\ell}  + \eta (\delta_{ik}\delta_{j\ell} + \delta_{i\ell}\delta_{jk} - \delta_{ij}\delta_{k\ell})\\
&+\eta_{\rm H} (\delta_{jk}\epsilon_{i\ell} - \delta_{i\ell}\epsilon_{kj})~.
\end{split}
\end{equation}
In Eq.~(\ref{eq:viscosity_tensor}), $\eta_{\rm H}$ parametrizes the portion of $\eta_{ij, k\ell}$ which is antisymmetric with respect to the exchange $ij \leftrightarrow k\ell$ and is non-zero only when time-reversal symmetry is broken.

\begin{figure}[t]
\centering
\begin{overpic}[width=.9\columnwidth]{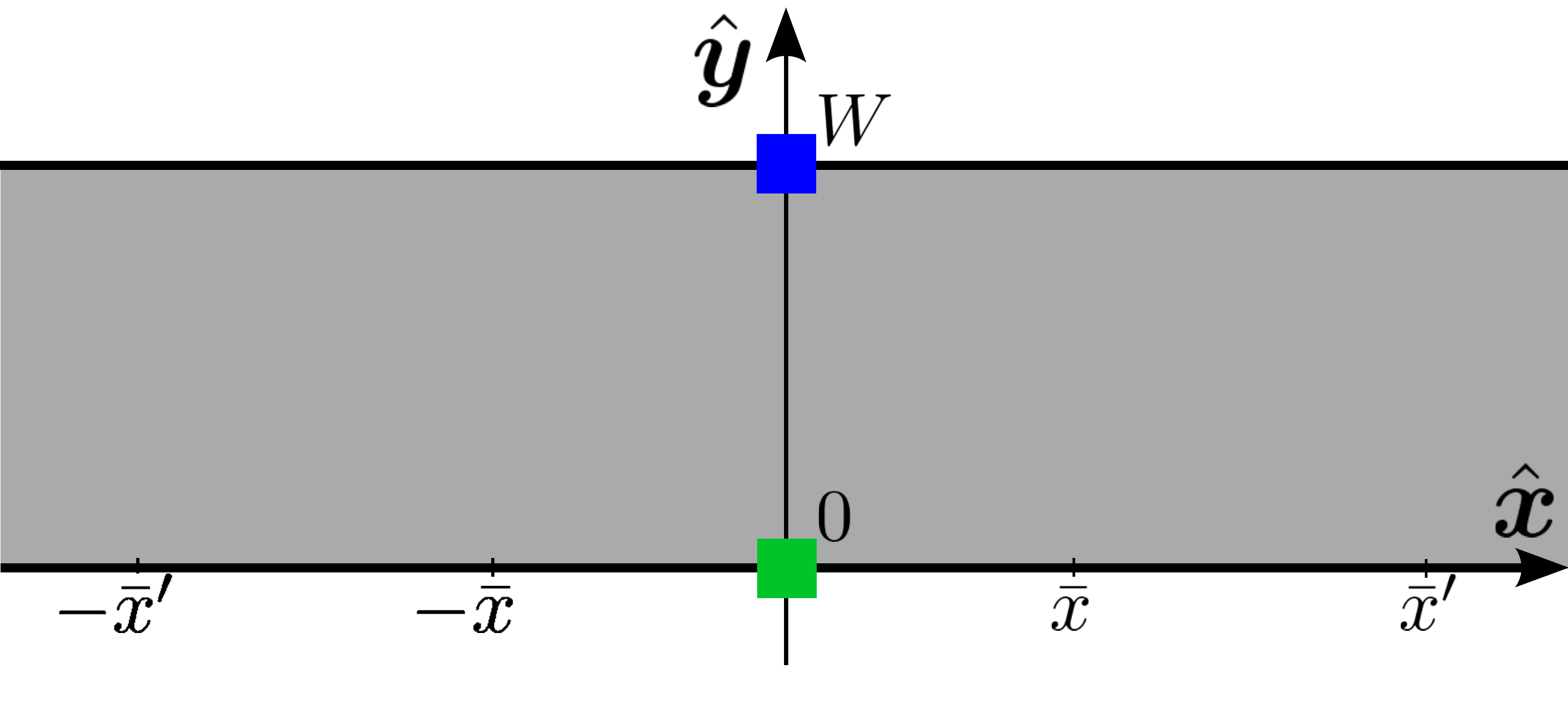}\put(2,38){(a)}
\end{overpic}\\
\begin{overpic}[width=.9\columnwidth]{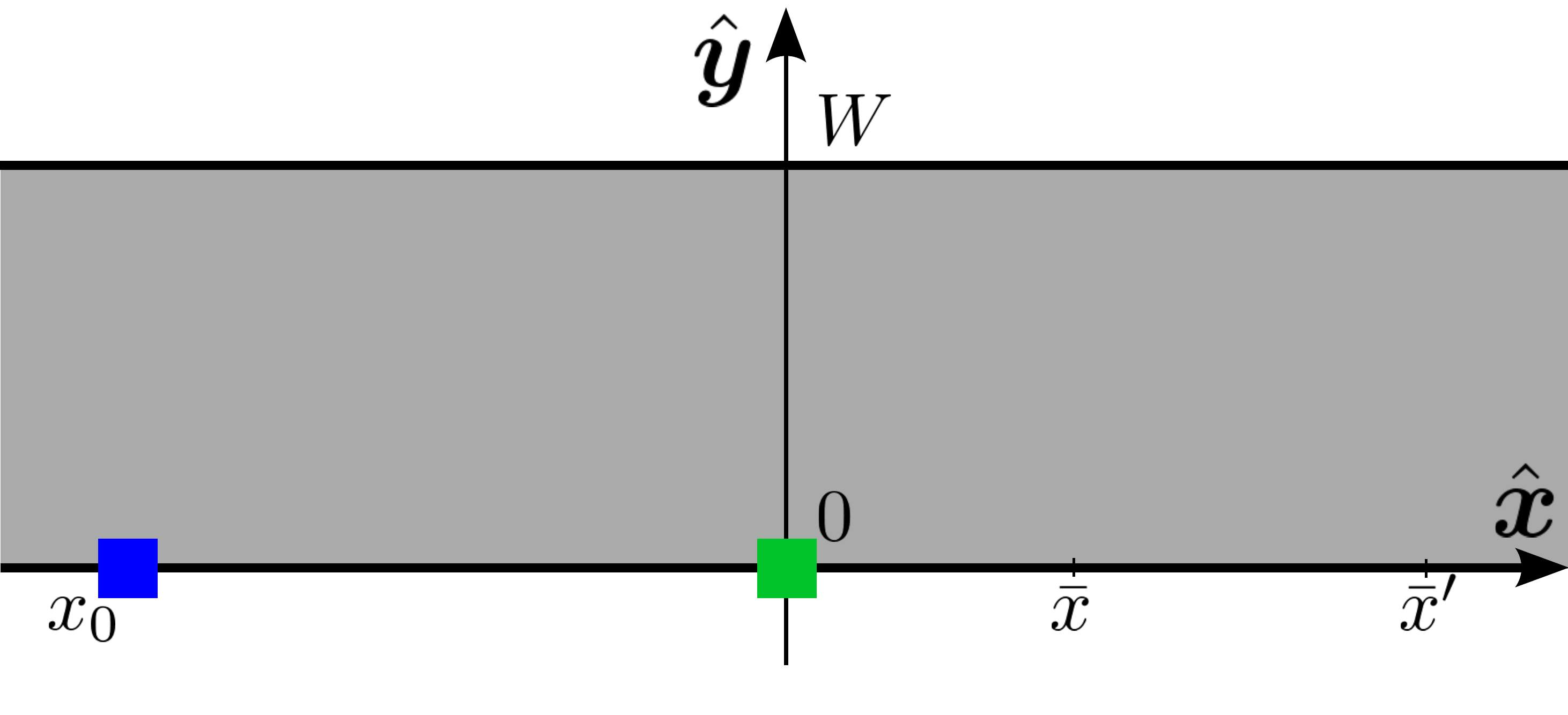}\put(2,38){(b)}
\end{overpic}
\caption{(Color online) A sketch of the non-local transport setups analyzed in this work. 
All setups have infinite length in the ${\hat {\bm x}}$ direction and finite width $W$ in the ${\hat {\bm y}}$ direction. 
Panel (a) illustrates the ``transverse'' geometry~\cite{levitov_naturephys_2016,pellegrino_prb_2016}. 
In this setup, current is injected into (extracted from) the green (blue) electrode located at $x=0$, $y = 0$ ($x=0$, $y = W$). 
Panel (b) illustrates the ``vicinity'' geometry~\cite{torre_prb_2015,pellegrino_prb_2016,bandurin_science_2016}. 
In this setup, current is injected into (extracted from) the green (blue) electrode located at $x=0$ ($x=x_{0}<0$) and $y=0$. 
\label{fig:setup}}
\end{figure}

Recent theoretical and experimental work has attracted interest in the flow of viscous electron liquids. In particular, two experiments~\cite{bandurin_science_2016,kumar_arxiv_2017} in high-quality encapsulated graphene sheets have demonstrated two unique qualitative features of viscous electron transport (negative quasi-local resistance~\cite{bandurin_science_2016} and super-ballistic electron flow~\cite{kumar_arxiv_2017}), providing, for the first time, the ability to directly measure the dissipative shear viscosity $\eta$ of a two-dimensional (2D) electron system. A different experiment~\cite{crossno_science_2016} has shown that near charge neutrality, electron-electron interactions in graphene are strong enough to yield substantial violations of the Wiedemann-Franz law. Evidence of hydrodynamic transport has also been reported in quasi-2D channels of palladium cobaltate~\cite{moll_science_2016}. Interestingly, also hydrodynamic features of phonon transport have been recently 
discussed in the literature~\cite{fugallo_nanolett_2014, capellotti_natcomm_2015}.

In this work, we focus on the role of the Hall viscosity $\eta_{\rm H}$ on the non-local electrical transport characteristics of 2D electron systems subject to a perpendicular magnetic field. Solving suitable magneto-hydrodynamic equations for the rectangular geometries sketched in Fig.~\ref{fig:setup}, we demonstrate how one can directly measure $\eta_{\rm H}$ by purely electrical non-local measurements.

Our Article is organized as follows. In Sect.~\ref{sec:magnetohydro} we review the theory of magneto-hydrodynamic transport in viscous 2D electron systems. 
In Sect.~\ref{sec:single} we present the solution of the magneto-hydrodynamic equations in the case of a rectangular setup, with infinite length in the $\hat{\bm x}$ direction and finite width in the $\hat{\bm y}$ direction, in the presence of  a single current injector on one side of the setup.
This solution is then used in Sect.~\ref{sec:RTV} as a building block to construct the solutions for the ``transverse'' and ``vicinity'' geometries, sketched in Figs.~\ref{fig:setup}~(a) and~(b), respectively. Finally, in Sect.~\ref{sec:conclusions} we summarize our principal findings and draw our main conclusions.
Detailed microscopic derivations of the magneto-hydrodynamic equations and of the corresponding boundary conditions are reported in Appendices~\ref{app:boltzmann} and~\ref{app:boundary}, respectively.
Appendix~\ref{app:noslip} discusses how the solution of Eqs.~(\ref{eq:continuity})-(\ref{eq:NavierStokes}) changes due to a change in the boundary conditions.

\section{Magneto-hydrodynamic theory of non-local transport}
\label{sec:magnetohydro}

In the linear-response and steady-state regimes, electron transport in the hydrodynamic regime in the presence of a static magnetic field $\bB= B \hat{\bm z}$ is described by the continuity equation
\begin{equation}\label{eq:continuity}
\nabla \cdot \bJ (\br)=0~,
\end{equation}
and the Navier-Stokes equation
\begin{equation}\label{eq:NavierStokes}
- \nabla P(\br) + \nabla \cdot {\bm \sigma}^\prime(\br) + e \bar{n} \nabla \varphi(\br) - \frac{e}{c} \bJ(\br) \times \bB=\frac{m }{\tau}\bJ(\br)~.
\end{equation}
Here,  $\bJ(\br)=\bar{n} \bv (\br)$ is the particle current density, $\bv (\br)$ is the fluid element velocity, $\bar{n}$ is the ground-state uniform density, 
 $P(\br)$ is the pressure, $\bm{\sigma}^\prime(\br)$ is the viscous stress tensor whose Cartesian components have been explicitly reported in Eqs.~(\ref{eq:stress_Cartesian_indices})-(\ref{eq:viscosity_tensor}), $\varphi(\br)$ is the 2D electrostatic potential in the plane where electrons move, $- e$ is the electron charge, $m$ is the electron effective mass, and
 $\tau$ is a phenomenological transport time describing momentum-non-conserving collisions~\cite{torre_prb_2015} (e.g.~scattering of electrons against acoustic phonons). The gradient of the pressure is proportional to the gradient of the density via $\nabla P(\br)=({\cal B}/\bar{n})\nabla n(\br)$,  
where ${\cal B}=\bar{n}^2/{\cal N}_0$ is the bulk modulus~\cite{Giuliani_and_Vignale} of the homogeneous electron liquid, ${\cal N}_0$ being the density of states at the Fermi energy~\cite{Giuliani_and_Vignale}. It is useful to define the electrochemical potential as $\phi(\bm r)=\varphi(\bm r)+ \delta \mu (\bm r)/(-e)$ where $\delta \mu(\bm r)=[n(\bm r)-\bar{n}]/{\cal N}_0$ is the chemical potential measured with respect to the equilibrium value, e.g.~$\bar{\mu}=\hbar v_{\rm F} \sqrt{\pi \bar{n}}$ for the case of single-layer graphene~\cite{kotov_rmp_2012} and $\bar{\mu}=\hbar^2 \pi \bar{n}/(2m)$ for bilayer graphene~\cite{kotov_rmp_2012}. 
Since experimental probes are usually sensitive to $\phi(\bm r)$, from now on we will focus our attention on the electrochemical potential rather than on $\varphi(\bm r)$.

We now note that the viscous stress tensor in Eqs.~(\ref{eq:stress_Cartesian_indices})-(\ref{eq:viscosity_tensor}) can be written in the following compact form
\begin{eqnarray}\label{eq:stress_complete}
\bm{\sigma}^\prime &=& (\eta +i \eta_{\rm H} \tau_{y})[(\partial_x v_x-\partial_y v_y)\tau_{z}+(\partial_x v_y+ \partial_y v_x)\tau_{x}]
\nonumber\\
&+& \zeta \nabla \cdot {\bm v}~,
\end{eqnarray}
where $\tau_{i}$ with $i=x,y,z$ are standard $2\times 2$ Pauli matrices acting on Cartesian indices. As in Eq.~(\ref{eq:NavierStokes}) above, in the linear-response and steady-state regimes we can write ${\bm v}({\bm r}) = {\bm J}({\bm r})/{\bar n}$. We then note that the bulk viscosity $\zeta$ couples to $\nabla \cdot {\bm J}$, which vanishes because of the continuity equation (\ref{eq:continuity}). The bulk viscosity term in the viscous stress tensor therefore drops out of the problem at hand. In summary, Eq.~(\ref{eq:stress_complete}) simplifies to:
\begin{equation}\label{eq:stress}
\bm{\sigma}^\prime=m (\nu +i \nu_{\rm H} \tau_{y})[(\partial_x J_x-\partial_y J_y)\tau_{z}+(\partial_x J_y+ \partial_y J_x)\tau_{x}]~,
\end{equation}
where $\nu \equiv \eta/(m \bar{n})$ is the kinetic shear viscosity and $\nu_{\rm H}\equiv \eta_{\rm H}/(m \bar{n})$ is the kinetic Hall viscosity. Replacing Eq.~(\ref{eq:stress}) into Eq.~(\ref{eq:NavierStokes}) and introducing the electrochemical potential $\phi({\bm r})$, we can write the Navier-Stokes equation (\ref{eq:NavierStokes}) as
\begin{equation}\label{eq:NS}
\frac{\sigma_0}{e} \nabla \phi(\br)=(1-D_\nu^2 \nabla^2)\bJ(\br) + \omega_{\rm c}\tau\left(1+D_{\rm H}^2   \nabla^2 \right) \bJ(\br) \times \hat{\bm z}~,
\end{equation}
where $\sigma_{0} =  n e^2\tau/m$, 
$D_{\nu}\equiv \sqrt{\nu \tau}$ has been introduced in Refs.~\onlinecite{torre_prb_2015,bandurin_science_2016,pellegrino_prb_2016},
$D_{\rm H}\equiv\sqrt{-\nu_{\rm H}/\omega_{\rm c}}$, and $\omega_{\rm c} \equiv e B/(mc)$ is the usual cyclotron frequency. As we will see below, $\nu_{\rm H}$ and $\omega_{\rm c}$ have opposite signs so that $D_{\rm H}$ is a well defined length scale. Notice that the Hall viscosity parametrizes a correction to the ordinary Lorentz force due to the spatial dependence of the velocity ${\bm v}({\bm r})$.

We now resort to useful results of semiclassical Boltzmann transport theory (see Appendix~\ref{app:boltzmann}), which capture the dependence of the shear and Hall viscosities on the magnitude of the applied magnetic field~\cite{steinberg_pr_1958}:
\begin{align}
\nu&=\nu_0 \frac{B_0^2}{B_0^2+B^2}\label{eq:nuB}\\
\nu_{\rm H}&=-\nu_0 \frac{B B_0}{B_0^2+B^2}\label{eq:nuHB}~,
\end{align}
where $\nu_0$ is the kinematic shear viscosity at zero magnetic field and $B_{0} \equiv c \bar{n}/(4 e {\cal N}_0 \nu_0)$ 
is a characteristic magnetic field. 
For example, for bilayer graphene~\cite{kotov_rmp_2012} with carrier density $\bar{n}=10^{12}~{\rm cm^{-2}}$,
we find $B_{0}\approx 0.1~{\rm Tesla}$ for~\cite{bandurin_science_2016,kumar_arxiv_2017} $\nu_0 \approx 0.1~{\rm m^2/s}$. For the same set of parameters and $|B| \ll B_0$, we find $D_{\rm H} \approx 1~\mu{\rm m}$.
Despite Eq.~(\ref{eq:nuHB}) has been derived in the weak-field limit, it yields sensible results even for high magnetic fields. In the limit $|B| \gg B_{0}$, indeed, we find $\nu_{\rm H} \approx -\sign(B)\ell_{B}^2 \bar{n}/(4 \hbar {\cal N}_0 )$, $\ell_B=\sqrt{c \hbar/(e|B|)}$ being the magnetic length, in agreement with the quantum Hall regime results derived in Ref.~\onlinecite{sherafati_prb_2016}.

Since all the setups in Fig.~\ref{fig:setup} are translationally-invariant in the ${\hat {\bm x}}$ direction, 
it is useful to introduce the following Fourier Transforms~\cite{torre_prb_2015,pellegrino_prb_2016} (FTs) with respect to the spatial coordinate $x$:
$\tilde{\phi}(k,y) = \int_{-\infty}^{+\infty}d x~e^{-i k x} \phi(\br)$
and
$\tilde{\bm J}(k,y) = \int_{-\infty}^{+\infty}d x~e^{-i k x} {\bm J}(\br)$.
The three coupled partial-differential equations (\ref{eq:continuity})-(\ref{eq:NavierStokes}) can be combined into a $4\times4$ system of first-order ordinary differential equations:
\begin{equation}\label{eq:M-matrix}
\partial_y{\bm w}(k,y) = {\cal M}(k) {\bm w}(k,y)~,
\end{equation}
where ${\bm w}(k,y)$ is a four-component vector, i.e. ${\bm w}(k,y)=[k \tilde{J}_x(k,y),k \tilde{J}_y(k,y),\partial_y\tilde{J}_x(k,y),e \bar{n} \tilde{\phi}(k,y)/(m \nu)]^{\rm T}$, and
\begin{widetext}
\begin{equation}\label{eq:Mmatrix}
{\cal M}(k) = k \left(
 \begin{array}{cccc}
0 & 0 & 1  & 0  \\
-i & 0 & 0 & 0 \\
1+1/(k D_{\nu})^2 & \nu_{\rm r} +\omega_{\rm c} \tau/(k D_\nu )^2  &i \nu_{\rm r} & -i\\
(\nu_{\rm r}-\omega_{\rm c} \tau) /(k D_{\nu})^2 &1 + \nu_{\rm r}^2 +(1 +\nu_{\rm r} \omega_{\rm c}\tau)/(k D_\nu)^2 &  i(1+\nu_{\rm r}^2) &-i \nu_{\rm r}
\end{array}
\right)~,
\end{equation}
\end{widetext}
where $\nu_{\rm r}\equiv\nu_{\rm H}/\nu$. 
The matrix ${\cal M}(k)$ has four eigenvalues: $\lambda_{1/2}(k) = \pm |k|$ and $\lambda_{3/4}(k) = \pm q$, where we have introduced the shorthand
\begin{equation}
 q \equiv \sqrt{k^2+1/D_\nu^2}~.
\end{equation}
The corresponding eigenvectors are:
\begin{equation}\label{eq:eigenvectors}
{\bm w}_{1/2}(k)=
\begin{pmatrix}
i\\
\pm  \sign(k)\\
\pm i \sign(k)\\
\frac{1 \mp i \sign(k) \omega_{\rm c} \tau}{D_\nu^2 k^2 }
\end{pmatrix}
,~
{\bm w}_{3/4}(k)=
\begin{pmatrix}
\pm \frac{k}{q}\\
- i \frac{k^2}{q^2}\\
1\\
\frac{(\nu_{\rm r} -\omega_{\rm c} \tau)}{D_\nu^2 q^2}
\end{pmatrix}~.
\end{equation}
Note that the eigenvalues are independent of the cyclotron frequency and Hall viscosity, while the eigenvectors explicitly depend on them.
The general solution of Eq.~(\ref{eq:M-matrix}) can be therefore written as a linear combination of exponentials of the form $\sum_{j=1}^4 a_j(k){\bm w}_j(k)\exp(\lambda_j  y)$. 
The four coefficients $a_j(k)$ can be determined from the enforcement of suitable boundary conditions (BCs).
\begin{figure}[t]
\centering
\begin{overpic}[width=\columnwidth]{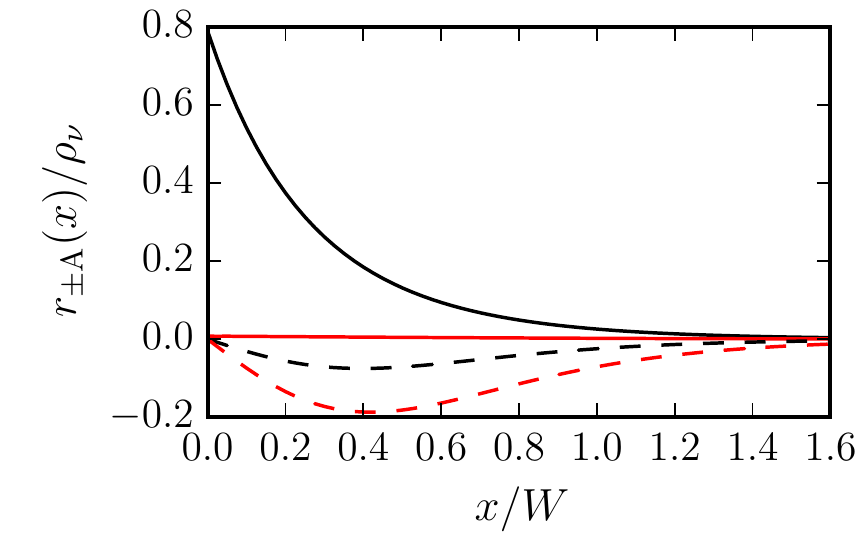}\put(2,58){(a)}\end{overpic}\\
\begin{overpic}[width=\columnwidth]{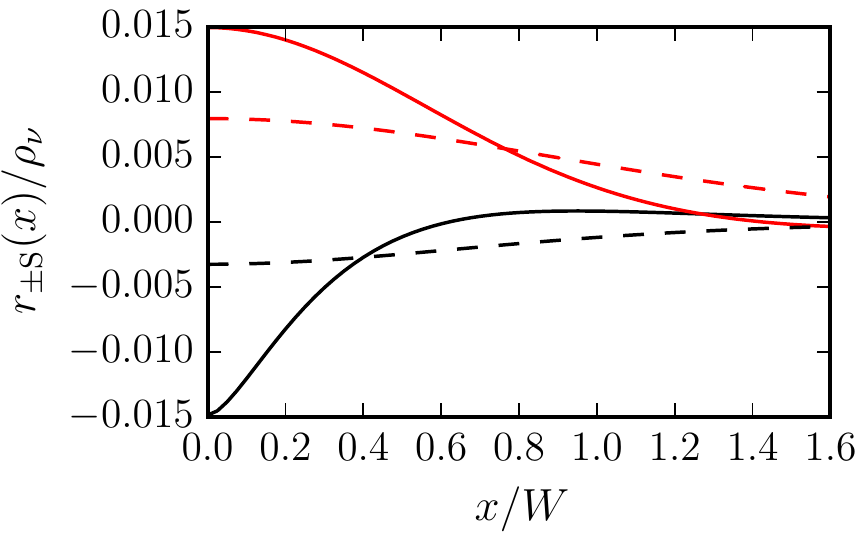}\put(2,58){(b)}\end{overpic}
\caption{(Color online) Panel (a) [Panel (b)] shows the inverse Fourier Transform of $\tilde{r}_{\rm \pm A}(k)$ [$\tilde{r}_{\rm \pm S}(k)$] in units of $\rho_\nu=m W^2/(\bar{n} e^2 \nu)$ and plotted as a function of $x/W$.
Results in both panels refer to a bilayer graphene sample with $W=2.5~\mu{\rm m}$, $\bar{n}=10^{12}~{\rm cm}^{-2}$, $B=0.1~B_{0}$, $\nu_0=0.1~{\rm m}^{2}/{\rm s}$, and free-surface BCs ($\ell_{\rm b}=\infty$). 
Different colors refer to different values of $\tau$.  Black: $\tau=2~{\rm ps}$. Red: $\tau=200~{\rm ps}$.
Solid lines refer to $\tilde{r}_{\rm + A}(x)$ and $\tilde{r}_{\rm + S}(x)$. Dashed lines to $\tilde{r}_{\rm - A}(x)$ and $\tilde{r}_{\rm - S}(x)$.\label{fig:rpm}}
\end{figure}
\section{Single-injector setup}
\label{sec:single}

We consider a single current injector in a rectangular setup with infinite length in the ${\hat {\bm x}}$ direction and finite width $W$ in the ${\hat {\bm y}}$ direction. This plays the role of ``building block'', allowing us to solve the magneto-hydrodynamic problem posed by Eqs.~(\ref{eq:continuity})-(\ref{eq:NavierStokes}) in more complicated setups like the ones sketched in Figs.~\ref{fig:setup}(a) and~(b).
 
A current injector is mathematically described by the usual point-like BC for the component of the current density perpendicular to the $y=0$ edge:
\begin{equation}
 J_y(x,0)= -  I\delta(x)/e~, 
\end{equation}
where $I$ in the dc drive current~\cite{abanin_science_2011}. 
On the edge opposite to the injector (i.e.~at $y=W$), the orthogonal component of the current density must vanish: 
\begin{equation}
 J_y(x, W)= 0~.
\end{equation}
The solution of the viscous problem requires additional BCs on the tangential components of the current density at both edges. We impose that the current density on the boundary $\Omega$ of the sample is proportional to the off-diagonal component of the stress~\cite{torre_prb_2015,pellegrino_prb_2016}:
\begin{equation}\label{eq:NavierBC}
[\hat{\bm e}_{\rm t} \cdot (\hat{\bm \sigma}^\prime \cdot \hat{\bm e}_{\rm n} ) + (m \nu/\ell_{\rm b}) \hat{\bm e}_{\rm t} \cdot \bJ ]_{\Omega}=0~,
\end{equation}
where $\hat{\bm e}_{\rm n}$ denotes the outer normal unit vector to the boundary, 
$\hat{\bm e}_{\rm t}=\hat{\bm e}_{\rm n}\times \hat{\bm z} $, and $\ell_{\rm b}$ is the so-called ``boundary slip length''. Using Boltzmann equation and the well-known Reuter-Sondheimer model of boundary scattering~\cite{reuter_rspa_1948}, we obtain (Appendix~\ref{app:boundary})
\begin{equation}\label{eq:lbBoltzmann}
\ell_{\rm b}=\frac{\nu}{v_{\rm F}}\frac{6\pi}{9\pi^2-32}\frac{(1+p)}{(1-p)}~,
\end{equation}
where $0\leq p \leq 1$ is the probability of specular scattering for an electron at the boundary~\cite{reuter_rspa_1948}. Here, $p=1$ for perfectly specular scattering and $p=0$ for completely diffusive scattering. Eq.~(\ref{eq:lbBoltzmann}) shows that $\ell_{\rm b}$ depends on both electron-electron scattering, through $\nu$, and electron-boundary scattering, through $p$. The boundary slip length diverges for $p\to 1$, recovering the free-surface BC~\cite{torre_prb_2015,pellegrino_prb_2016}, 
while it remains finite in the limit $p\to 0$, i.e.~$\lim_{p\to 0}\ell_{\rm b} \simeq 0.1~\ell_{\rm ee}$, where $\ell_{\rm ee}$ is the electron-electron scattering length~\cite{principi_prb_2016}. (In deriving the last result we have used $\nu \simeq v_{\rm F}\ell_{\rm ee}/4$, see Appendix~\ref{app:boltzmann}.) For completely diffusive scattering, $\ell_{\rm b}$ is therefore ten times smaller than the electron-electron scattering length. Since the latter quantity is much smaller than the macroscopic length scales of hydrodynamic electron flow, $\ell_{\rm b}$ is negligibly small for $p\to 0$ and we obtain the no-slip BC~\cite{torre_prb_2015,levitov_naturephys_2016}. Although Eq.~(\ref{eq:lbBoltzmann}) has been obtained by using a very simple model for boundary scattering~\cite{reuter_rspa_1948}, we believe that more refined models will yield different numerical values for $\ell_{\rm b}$ but will likely not change neither the structure of Eq.~(\ref{eq:NavierBC}) nor the qualitative conclusions we just drew.

\begin{figure}[t]
\centering
\begin{overpic}[width=\columnwidth]{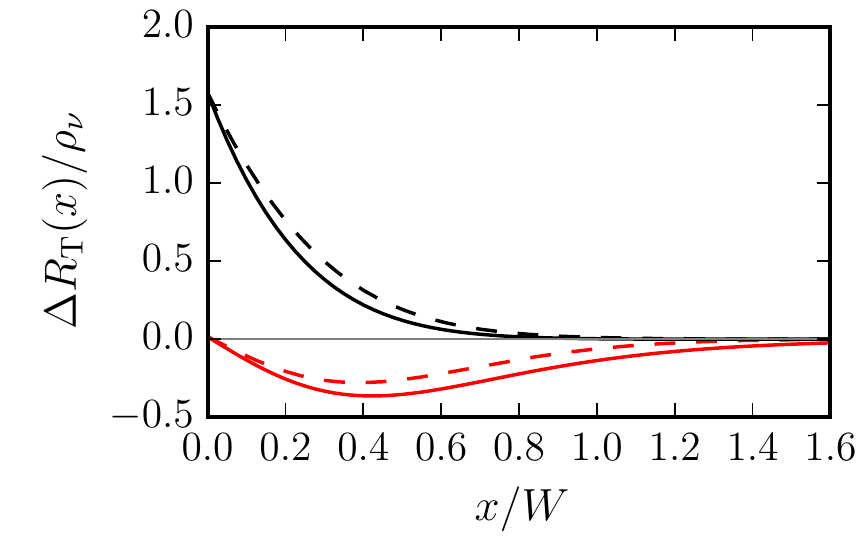}\put(2,58){(a)}\end{overpic}\\
\begin{overpic}[width=\columnwidth]{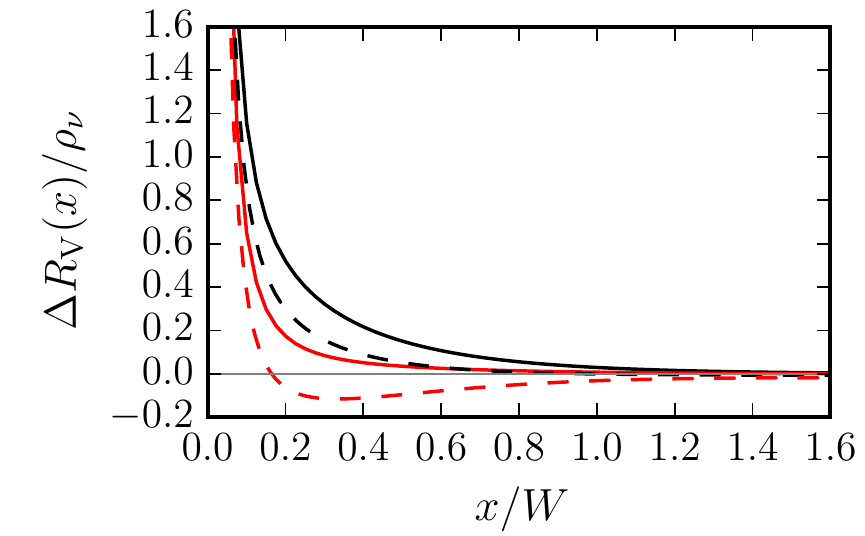}\put(2,58){(b)}\end{overpic}
\caption{(Color online)  Panel (a) The transverse  resistance difference (\ref{eq:DRT}), in units of $\rho_{\nu}$, plotted as a function of $x/W$.
Panel (b) The vicinity  resistance difference (\ref{eq:DRV}), in units of $\rho_{\nu}$, plotted as a function of $x/W$.
Results in both panels refer to a bilayer graphene sample with $W=2.5~\mu{\rm m}$, $\bar{n}=10^{12}~{\rm cm}^{-2}$, $B=0.1~B_{0}$, and $\nu_0=0.1~{\rm m}^{2}/{\rm s}$. 
Different colors refer to different values of $\tau$.  Black: $\tau=2~{\rm ps}$. Red: $\tau=200~{\rm ps}$.
Solid lines refer to $\ell_{\rm b}\to \infty$ (free-surface BCs). Dashed lines to $\ell_{\rm b}\to 0$ (no-slip BCs).
\label{fig:Rx}}
\end{figure}

For the setups in Fig.~\ref{fig:setup}, we have $\hat{\bm e}_{\rm n} =  \hat{\bm y}$ 
($\hat{\bm e}_{\rm n} = - \hat{\bm y} $) for the upper (lower) edge at $y=W$ ($y=0$), respectively. In FT with respect to $x$, the BCs become
\begin{align}
& \partial_y  \tilde{J}_{x}(k,0)+i k \tilde{J}_y(k,0)  =(2 i k \nu_{\rm r} + \ell_{\rm b}^{-1})\tilde{J}_{x}(k,0)~,\\
& \tilde{J}_y(k,  0)=- I /e~,\\ 
& \partial_y \tilde{J}_{x}(k,W)=(2 i k \nu_{\rm r} - \ell_{\rm b}^{-1})\tilde{J}_{x}(k,W)~,\\
& \tilde{J}_y(k, W) =0~. 
\end{align}
In the remainder of this Section, we consider the case of free-surface BCs~\cite{torre_prb_2015,pellegrino_prb_2016}, which are obtained by taking the limit $\ell_{\rm b} \to +\infty$. 
This choice is physically justified by the measured~\cite{bandurin_science_2016} {\it monotonic} temperature dependence (i.e.~no Gurzhi effect) of the ordinary longitudinal resistance in the linear-response regime and in the case of a uniform steady-state flow. For more details, we refer the reader to Refs.~\onlinecite{torre_prb_2015,pellegrino_prb_2016,bandurin_science_2016}.
To study the impact of a boundary slip length $\ell_{\rm b}< +\infty$, we have carried out the same calculations described in this Section in the opposite limit, i.e.~for~$\ell_{\rm b}=0$ (See Appendix~\ref{app:noslip}).
The results obtained for $\ell_{\rm b} \to +\infty$ and $\ell_{\rm b} = 0$ are compared in Sect.~\ref{sec:RTV}, see Fig.~\ref{fig:Rx}.

The FT of the electrochemical potential along the edges reads as following:
\begin{widetext}
\begin{equation}\label{eq:phi+}
\begin{split}
\tilde{\phi}_+ (k) & \equiv \tilde{\phi}(k,0)=\\
&=\frac{I \rho_0}{ k }\Big\{ \sinh (\bar{q}) [(1 + 2 D_\nu^2 k^2) \cosh (\bar{k})+ i \omega_{\rm c} \tau  \sinh (\bar{k})]+i 4  \nu_{\rm H} \nu \tau^2  k^2 q  \{q \sinh (\bar{k}) \sinh (\bar{q})-k[\cosh (\bar{k}) \cosh (\bar{q})-1]\}\\
&+i2  \nu_{\rm H}^2 \tau^2 k^2  \big \{2 k q (2 \omega_{\rm c} \tau-\nu_{\rm r}) [1-\cosh (\bar{k}) \cosh (\bar{q})]-\sinh (\bar{q})\{iD_\nu^{-2}\cosh (\bar{k})-2 \sinh (\bar{k}) [k^2 (\omega_{\rm c} \tau-\nu_{\rm r})+ q^2 \omega_{\rm c} \tau]  \} \big\} \Big\}\\
&\times \Big\{\sinh (\bar{k}) \sinh (\bar{q}) +4  \nu_{\rm H}^2 \tau^2 k^2 \{(2k^2+D_\nu^{-2}) \sinh (\bar{k}) \sinh (\bar{q})+2 k q[1- \cosh (\bar{k}) \cosh (\bar{q})]\} \Big\}^{-1},
\end{split}
\end{equation}
\begin{equation}\label{eq:phi-}
\begin{split}
\tilde{\phi}_-(k) \equiv \tilde{\phi}(k,W) & = \frac{I\rho_0}{ k}\Big\{(1 + 2 D_\nu^2  k^2) \sinh(\bar{q})+2\nu_{\rm H} \tau k^2  \{ \nu_{\rm r} \sinh (\bar{q})+2 i k q D_\nu^{2}(1+\nu_{\rm r}^2) [\cosh (\bar{k})-\cosh (\bar{q})]\}\Big\}\\   
&\times \Big\{\sinh (\bar{k}) \sinh (\bar{q}) +4  \nu_{\rm H}^2 \tau^2 k^2 \{(2k^2+D_\nu^{-2}) \sinh (\bar{k}) \sinh (\bar{q})+2 k q[1- \cosh (\bar{k}) \cosh(\bar{q})]\} \Big\}^{-1},
\end{split}
\end{equation}
\end{widetext}
where $\bar{k}=kW$, $\bar{q}=qW$, $\rho_0=\sigma_0^{-1}$, and $\sigma_{0}=\bar{n}e^2 \tau/m$ represents a Drude-like conductivity. 
It is useful to express the edge electrochemical potentials (\ref{eq:phi+})-(\ref{eq:phi-}) as
\begin{equation}\label{eq:phip}
\tilde{\phi}_+(k) = I[\tilde{r}_+(k)- i\rho_{\rm H}/k+ 2i \rho_{\nu_{\rm H}} k W^2 +\tilde{r}_{\rm + S}(k)+ \tilde{r}_{\rm + A}(k)]~,
\end{equation}
\begin{equation}\label{eq:phim}
\tilde{\phi}_-(k) = I[\tilde{r}_-(k)+\tilde{r}_{\rm - S}(k)+ \tilde{r}_{\rm - A}(k)]~,
\end{equation}
where $\rho_{\rm H}=-m \omega_{\rm c}/(\bar{n} e^2)=B/(-e\bar{n}c)$, and $\rho_{\nu_{\rm H}}=m \nu_{\rm H}/(\bar{n} e^2 W^2)=\rho_{\rm H} D_{\rm H}^2/W^2$. 
The former quantity is the usual Hall resistivity.
The resistances $\tilde{r}_{\pm} (k)$ physically represent the solutions at zero magnetic field~\cite{torre_prb_2015,pellegrino_prb_2016} and are given by
\begin{equation}\label{eq:r+k}
\tilde{r}_{+}(k)=\frac{\rho_0 + 2 \rho_\nu k^2 W^2 }{k\tanh(k W)}
\end{equation}
and 
\begin{equation}\label{eq:r-k}
\tilde{r}_-(k)=\frac{\rho_{0} + 2 \rho_\nu k^2 W^2  }{\sinh(k W)k}~,
\end{equation}
where  $\rho_{\nu}=m \nu/(\bar{n} e^2 W^2)$.
The inverse FT of $\tilde{r}_{+}(k)$ and $\tilde{r}_{+}(-)$ can be calculated analytically. We find
\begin{equation}
r_+(x)=-\frac{\rho_0}{2 \pi} \ln \Big [\sinh^2\Big(\frac{\pi x}{2W}\Big)\Big]-\frac{\pi \rho_\nu}{2\sinh^2\big(\frac{\pi x}{2W}\big)}
\end{equation}
and
\begin{equation}
r_-(x)=-\frac{\rho_0}{2 \pi} \ln \Big [\cosh^2\Big(\frac{\pi x}{2W}\Big)\Big]+\frac{\pi \rho_\nu}{2\cosh^2\big(\frac{\pi x}{2W}\big)}~.
\end{equation}
In Eqs.~(\ref{eq:phip}) and~(\ref{eq:phim}), the quantity $\tilde{r}_{\rm \pm S}(k)$ ($\tilde{r}_{\rm \pm A}(k)$) is a real (imaginary) function and it is non zero only for a finite value of the kinematic Hall viscosity $\nu_{\rm H}$.
 
Furthermore, $\tilde{r}_{\rm \pm S}(k)$ ($\tilde{r}_{\rm \pm A}(k)$) is even (odd) under the exchange $k \to -k$, implying that the corresponding inverse FTs are even (odd) real functions of the coordinate $x$.
At zero magnetic field, $\omega_{\rm c}$ and $\nu_{\rm H}$ vanish, implying that $\tilde{r}_{\rm \pm S}(k)=0$ and $\tilde{r}_{\rm \pm A}(k)=0$. 

After straightforward mathematical manipulations, we find:
\begin{equation}\label{eq:rp}
\begin{split}
\phi_+(x) & =I\bigg\{-\frac{\rho_0}{2 \pi} \ln \Big [\sinh^2\Big(\frac{\pi x}{2W}\Big)\Big]-\frac{\pi \rho_\nu}{2\sinh^2\big(\frac{\pi x}{2W}\big)}\\
& +\frac{\rho_{\rm H}}{2} \sign(x)+2 \rho_{\nu_{\rm H}} \delta^\prime\left(\frac{ x}{W}\right) + r_{\rm + S}(x) + r_{\rm + A}(x)  \bigg\}
\end{split}
\end{equation}
and
\begin{equation}\label{eq:rm}
\begin{split}
\phi_-(x)=I\bigg\{ & -\frac{\rho_0}{2 \pi} \ln \Big [\cosh^2\Big(\frac{\pi x}{2W}\Big)\Big]+\frac{\pi \rho_\nu}{2\cosh^2\big(\frac{\pi x}{2W}\big)}\\
& + r_{\rm - S}(x) + r_{\rm - A}(x) \bigg\}~,
\end{split}
\end{equation}
where 
the third term on the right hand side of Eq.~(\ref{eq:rp}) is the usual contribution due to the Lorentz force,
the fourth term in the same equation is a singular contribution (proportional to the derivative of the Dirac delta function) localized at the position of the injector and due to the Hall viscosity, while
$r_{\rm \pm S}(x)$ and $r_{\rm \pm A}(x) $ are the inverse FTs of $\tilde{r}_{\rm \pm S}(k)$  and $\tilde{r}_{\rm \pm A}(k)$, which must be calculated numerically. Fig.~\ref{fig:rpm}(a) (Fig.~\ref{fig:rpm}(b)) shows the resistance  $r_{\rm \pm A}(x)$ ($r_{\rm \pm S}(x)$),  in units of $\rho_\nu$, plotted as a function of $x/W$. These calculations refer to the massive ($m=0.03~m_{\rm e}$, where $m_{\rm e}$ is the bare electron mass in vacuum) chiral 2D electron system~\cite{kotov_rmp_2012} in a bilayer graphene sample with $W=2.5~\mu{\rm m}$, $\bar{n}=10^{12}~{\rm cm}^{-2}$, $B=0.1~B_{0}$, $\nu_0=0.1~{\rm m}^{2}/{\rm s}$, with the black lines referring to $\tau=2~{\rm ps}$ and the red lines to the ultra-clean limit, $\tau=200~{\rm ps}$. 
The quantities $r_{\rm + A}(x)$ and $r_{\rm + S}(x)$ are denoted by solid lines, while $r_{\rm - A}(x)$ and $r_{\rm - S}(x)$ are denoted by dashed lines.
For small magnetic fields and in the ultra-clean $\tau \to \infty$ limit we can linearize $r_{\rm \pm A}(x)$ with respect to $B$ and $1/\tau$, obtaining the following analytical expressions:
\begin{equation}
r_{\rm +A}(x)\approx\frac{\rho_{\rm \nu_{\rm H}} W^2}{4D_\nu^2} \Bigg[\coth\left( \frac{\pi x}{2 W}\right)-\sign(x) - \frac{ \frac{\pi x}{2 W}}{\sinh^2\left( \frac{\pi x}{2 W}\right)}  \Bigg]
\end{equation}
and
\begin{equation}
r_{\rm -A}(x)\approx\rho_{\rm \nu_{\rm H}}\Bigg[\frac{\pi^2 \tanh\left( \frac{\pi x}{2 W}\right)}{2 \cosh^2\left( \frac{\pi x}{2 W}\right)} -\frac{\pi W x}{8D_\nu^2  \cosh^2\left( \frac{\pi x}{2 W}\right) } \Bigg]~,
\end{equation}
which are in excellent agreement with the numerical results shown for $\tau = 200~{\rm ps}$ in Fig.~\ref{fig:rpm}(a). In the limit $B/B_{0}\ll 1$, the quantities $\tilde{r}_{\rm \pm S}(k)$ start at order $(B/B_{0})^2$.

\section{Non-local resistances and the Hall viscosity}
\label{sec:RTV}

We now turn to discuss explicitly the impact of the Hall viscosity on non-local electrical transport measurements carried out in the two setups sketched in Fig.~\ref{fig:setup}(a) and (b). We start from the setup depicted in Fig.~\ref{fig:setup}(a), 
where the current $I$ is injected into the green electrode at $x=0$ and $y=0$, and extracted from the blue electrode at $x=0$ and $y=W$. 
As in the case of the single-injector setup we just discussed, we treat each electrode with the usual point-like BC for the component of current density orthogonal to the edge. 

Exploiting the linearity of the problem, it is possible to write the electrochemical potentials along the edges of the setup in Fig.~\ref{fig:setup}(a) in terms of the potentials $\phi_+(x)$ and $\phi_-(x)$ along the lower and upper edges of the single-injector setup, respectively: 
\begin{align}
\phi(x,0)&=\phi_+(x)-\phi_-(-x)~,\\ 
\phi(x,W)&=\phi_-(x)-\phi_+(-x)~. 
\end{align}
We remind the reader that $\phi_{+}(x)$ and $\phi_{-}(x)$ are the inverse FTs of $\tilde{\phi}_{+}(k)$ and $\tilde{\phi}_{-}(k)$, respectively. For the case of the free-surface BCs ($\ell_{\rm b} \to +\infty$), explicit expressions for the latter quantities have been given in Eqs.~(\ref{eq:phi+}) and~(\ref{eq:phi-}). For the no-slip BCs, we refer the reader to Eqs.~(\ref{eq:phip_repeated})-(\ref{eq:CR7}) in Appendix~\ref{app:noslip}.

We now introduce the ``transverse'' non-local resistance, which is measured in the setup sketched in Fig.~\ref{fig:setup}(a), as
\begin{equation}\label{eq:RT}
\begin{split}
R_{\rm T}(x)&\equiv \frac{\phi(x,0)-\phi(-x,0)}{I}= \rho_{\rm H} \sign(x)+4 \rho_{\nu_{\rm H}} \delta^\prime\Big(\frac{ x}{W}\Big)\\
&+2[r_{\rm +A}(x)+r_{\rm -A}(x)]~.
\end{split}
\end{equation}
We note that $R_{\rm T}(x) \to \rho_{\rm H} \sign(x) $ for $|x| \gg W$, because, in the same limit, $[r_{\rm +A}(x)+r_{\rm -A}(x)]\to 0$, independently of the value of $\ell_{\rm b}$. In order to have a clear signature of the Hall viscosity it is therefore convenient to perform two measurements of the transverse resistance $R_{\rm T}$, i.e.~one at position $0<x\lesssim W$ and a second one at position $x^\prime \gg W$. The difference 
\begin{equation}\label{eq:DRT}
\Delta R_{\rm T}(x)\equiv  R_{\rm T}(x)- \lim_{x^\prime \to \infty} R_{\rm T}(x^\prime)=2[r_{\rm +A}(x)+r_{\rm -A}(x)]
\end{equation}
is independent of $\rho_{\rm H}$ and non-zero only in the presence of a finite Hall viscosity. 
Results in the transverse geometry show a weak dependence on the BCs (\ref{eq:NavierBC}).
Fig.~\ref{fig:Rx}(a) shows the quantity $\Delta R_{\rm T}(x)$ as a function of $x/W$, as calculated by using the BCs (\ref{eq:NavierBC}) in the two limiting cases, $\ell_{\rm b} \to +\infty$  (solid lines) and $\ell_{\rm b} \to 0$ (dashed lines). For the calculations we have used two different values of $\tau$:
$\tau=2~{\rm ps}$ (black) and $\tau=200~{\rm ps}$ (red).
From Fig.~\ref{fig:Rx} we clearly see a weak dependence of $\Delta R_{\rm T}(x)$ on $\ell_{\rm b}$, independently of the value of $\tau$.

We now follow similar algebraic steps for the setup in Fig.~\ref{fig:setup}(b).
In this case the current $I$ is injected into the green electrode at $x=0$ and $y=0$, and extracted from the blue electrode at $x=x_{0}<0$ and $y=0$. We find
\begin{align}
\phi(x,0)&=\phi_+(x)-\phi_+(x-x_0)~,\\ 
\phi(x,W)&=\phi_-(x)-\phi_-(x-x_0)~. 
\end{align}
We define the ``vicinity'' resistance~\cite{torre_prb_2015,bandurin_science_2016,pellegrino_prb_2016}  as
\begin{equation}\label{eq:RV}
 R_{\rm V}(x)\equiv \frac{\phi(x,0)-\phi(x^\prime \to +\infty,0)}{I}~.
\end{equation}
The mathematical expression of $R_{\rm V}(x)$ notably simplifies in the limit $x_{0} \to -\infty$, becoming
\begin{equation}
\begin{split}
 R_{\rm V}(x)&=r_{+}(x)+\frac{\rho_{\ast} x}{2W}+r_{\rm +A}(x)+r_{\rm +S}(x)\\
 &-[r_{\rm +A}(+ \infty)+r_{\rm +S}(+ \infty)]~, 
 \end{split}
\end{equation}
where $x>0$ and the resistance $\rho_{\ast}$ is obtained by the asymptotic relation $\rho_{\ast}  =-2W \lim_{x\to \infty} r_{+}(x)/|x|$.
In the limit $\ell_{\rm b}\to \infty$ (i.e.~free-surface BCs) and using Eq.~(\ref{eq:rp}) we find $\rho_{\ast}=\rho_{0}$. 
In the opposite limit, $\ell_{\rm b}\to 0$ (i.e.~no-slip BCs), we find $\rho_{\ast}=\rho_0\{1-2 D_\nu/W\tanh[W/(2D_\nu)]\}^{-1}$ (see Appendix~\ref{app:noslip}).

Since we are interested in the impact of the Hall viscosity on  hydrodynamic electrical transport,
it is useful to concentrate our attention on the difference between the vicinity resistance in the presence of an applied magnetic field and  in the absence of it:
\begin{equation}\label{eq:DRV}
 \Delta R_{\rm V}(x)\equiv  R_{\rm V}(x)- R_{\rm V}(x)|_{B=0}~.
\end{equation}
The vicinity geometry displays a non-trivial dependence on the BCs (\ref{eq:NavierBC}).
In Fig.~\ref{fig:Rx}(b) we show the quantity $\Delta R_{\rm V}(x)$ as a function of $x/W$, as calculated by using the BCs (\ref{eq:NavierBC}) in the two limiting cases, $\ell_{\rm b} \to +\infty$  (solid lines) and $\ell_{\rm b} \to 0$ (dashed lines). As in panel (a) of the same figure, we have carried out calculations for two different values of $\tau$:
$\tau=2~{\rm ps}$ (black) and $\tau=200~{\rm ps}$ (red).
In the ultra-clean limit ($\tau = 200~{\rm ps}$) the dependence of $\Delta R_{\rm T}(x)$ on $\ell_{\rm b}$ is large. Indeed, by comparing the solutions with free-surface and no-slip BCs, we note from Fig.~\ref{fig:Rx}(b) that even the sign of $\Delta R_{\rm T}(x)$ depends on $\ell_{\rm b}$.

\begin{figure}[t]
\vspace{0.5em}
\centering
\begin{overpic}[width=\columnwidth]{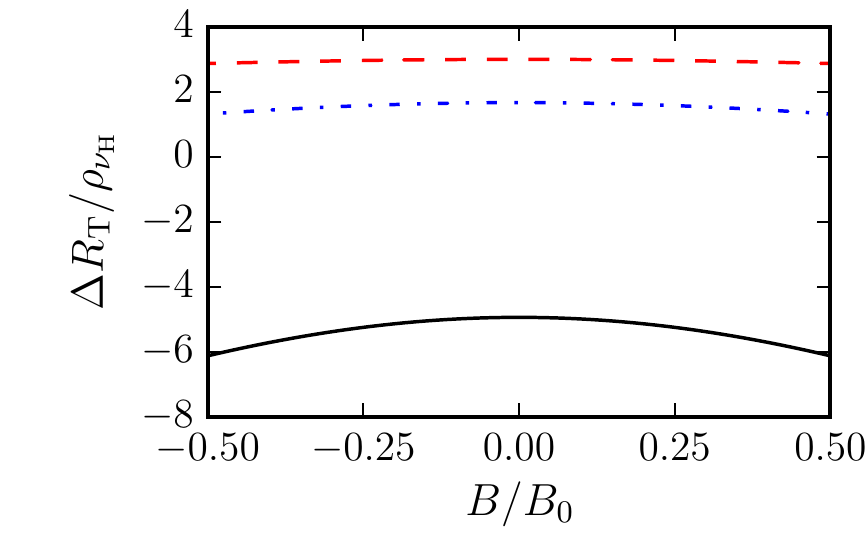}\put(2,58){(a)}\end{overpic}\\
\begin{overpic}[width=\columnwidth]{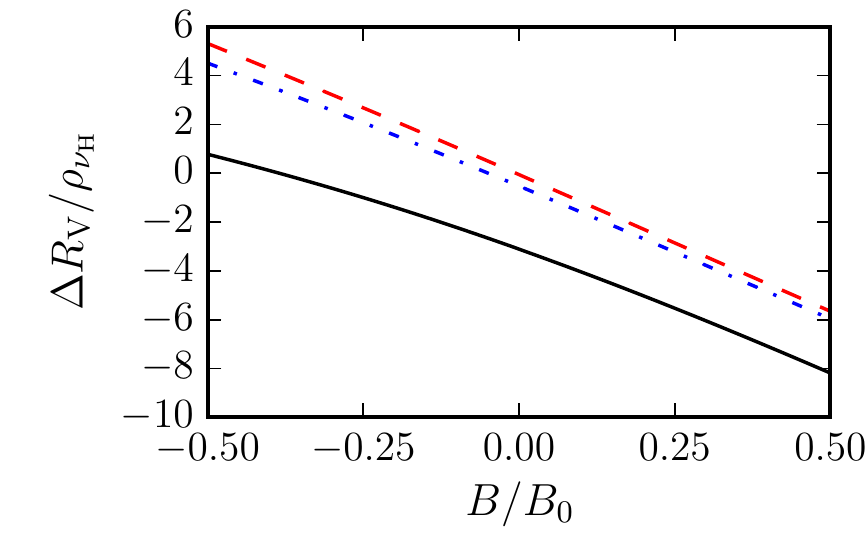}\put(2,58){(b)}\end{overpic}
\caption{(Color online) Panel (a) The transverse resistance difference (\ref{eq:DRT}), in units of $\rho_{\nu_{\rm H}}$ and evaluated at $x=0.25~W$, is plotted as a function of $B/B_{0}$. Panel (b) The vicinity resistance difference (\ref{eq:DRV}), in units of 
$\rho_{\nu_{\rm H}}$ and evaluated at $x=0.25~W$, is plotted as a function of $B/B_{0}$.
Results in both panels refer to a bilayer graphene sample with $W=2.5~\mu{\rm m}$, $\bar{n}=10^{12}~{\rm cm}^{-2}$, and $\nu_0=0.1~{\rm m}^{2}/{\rm s}$. All results shown in this figure have been evaluated by using the free-surface BCs ($\ell_{\rm b} = +\infty$). Solid line: $\tau=2~{\rm ps}$. Dash-dotted line: $\tau=20~{\rm ps}$. Dashed line: $\tau=200~{\rm ps}$.\label{fig:Res}}
\end{figure}

Before concluding, in Fig.~\ref{fig:Res}(a) and (b) we illustrate the dependence of 
$\Delta R_{\rm T}(x)$ and $\Delta R_{\rm V}(x)$ on $B/B_{0}$, respectively. In this case, these quantities have been calculated by using the free-surface BCs and evaluated at a given position $x \lesssim W$. In the weak-field $B\ll B_{0}$ limit, $\Delta R_{\rm T}$ is given by the product of $\rho_{\nu_{\rm H}}$ and a function that depends only on $x$, $\tau$, and $\nu(B=0)$. 
A measurement of $\Delta R_{\rm T}$ therefore yields immediately the value of the Hall viscosity, provided that $\tau$ is measured from the ordinary longitudinal resistance~\cite{bandurin_science_2016} $\rho_{xx}$ and $\nu(B=0)$ from one of the protocols discussed in Refs.~\onlinecite{bandurin_science_2016,kumar_arxiv_2017}. We emphasize that this way of accessing $\nu_{\rm H}$ is insensitive to the classical Hall resistivity $\rho_{\rm H}$ and to ballistic effects like transverse magnetic focusing~\cite{beconcini_prb_2016}. The latter statement holds true as long as $\nu_{\rm H}$ is extracted from a measurement of $\Delta R_{\rm T}$ at values of $B$ that 
are well below those that are necessary to focus electron trajectories~\cite{taychatanapat_naturephys_2013,beconcini_prb_2016}, for typical sample sizes. 

The vicinity geometry is in practice less convenient to probe $\nu_{\rm H}$. This is because---as seen in Fig.~\ref{fig:Res}(b)---the range of values of $B/B_{0}$ over which $\Delta R_{\rm V}$ depends on $B$ solely through the Hall viscosity is smaller than in the highly-symmetric geometry shown in Fig.~\ref{fig:setup}(a). We also mention once more that the geometry sketched in Fig.~\ref{fig:setup}(a) is more convenient for accessing $\nu_{\rm H}$ with respect to the one in Fig.~\ref{fig:setup}(b), because in the former one the detailed nature of BCs does not influence in a significant matter the role played by the Hall viscosity on non-local electrical measurements. In other words, in the case of Fig.~\ref{fig:setup}(a) and corresponding $\Delta R_{\rm T}(x)$, a precise estimate of $\ell_{\rm b}$ is unnecessary. This is at odds with a recently studied geometry~\cite{scaffidi_prl_2017}, 
where the impact of $\nu_{\rm H}$ on the electrochemical potential at the boundaries of the setup exists only for finite values of the boundary slip length ($\ell_{\rm b} < +\infty$). 
In our case, the electrochemical potential at the boundaries depends non-trivially on $\nu_{\rm H}$ for both free-surface ($\ell_{\rm b} = +\infty$) and no-slip ($\ell_{\rm b}=0$) BCs.

\section{Summary and conclusions}
\label{sec:conclusions}

In summary, we have proposed an all-electrical scheme that allows a determination of the Hall viscosity $\nu_{\rm H}$ of a two-dimensional electron liquid in a solid-state device. We have carried out extensive calculations for two device geometries, illustrated in Figs.~\ref{fig:setup}(a) and~(b), and a family of boundary conditions, reported in Eq.~(\ref{eq:NavierBC}), which depends on one parameter, the so-called boundary slip length $\ell_{\rm  b}$. The latter allows to interpolate between the widely used no-slip ($\ell_{\rm b}=0$) and free-surface ($\ell_{\rm b}=+\infty$) boundary conditions.

We have demonstrated that the transverse geometry in Fig.~\ref{fig:setup}(a) is particularly suitable for extracting $\nu_{\rm H}$ from experimental data. 
Indeed, we have shown that a measurement of $\Delta R_{\rm T}(x)$---Eq.~(\ref{eq:DRT})---yields immediately the value of the Hall viscosity, provided that $\tau$ is measured from the ordinary longitudinal resistance~\cite{bandurin_science_2016} $\rho_{xx}$ and $\nu(B=0)$ from one of the protocols discussed in Refs.~\onlinecite{bandurin_science_2016,kumar_arxiv_2017}. 
We have also shown that $\Delta R_{\rm T}(x)$ is insensitive to the value of the boundary slip length, a finding that further enforces the robustness of this quantity as a diagnostic tool of the Hall viscosity.

\noindent {\it Note added.---}While preparing this manuscript we became aware of related work~\cite{delacretaz_arxiv_2017} where the effect of the Hall viscosity on the dc flow of an electron fluid was studied by neglecting the impact of momentum-non-conserving collisions.

\begin{acknowledgments}
This work was supported by Fondazione Istituto Italiano di Tecnologia and the European Union's Horizon 2020 research and innovation programme under grant agreement No.~696656 ``GrapheneCore1''.
\end{acknowledgments}

\appendix

\section{Derivation of the linearized hydrodynamic equations from the semiclassical Boltzmann equation}
\label{app:boltzmann}
For sufficiently long-wavelength ($\lambda \gg 2\pi/k_{\rm F}$) and low-frequency ($\omega \ll 2E_{\rm F}/\hbar$, where $E_{\rm F}$ is the Fermi energy) perturbations, the response of a 2D electron system can be described by using the semiclassical Boltzmann equation~\cite{katsnelson}:
\begin{equation}\label{eq:Boltzmann}
\begin{split}
\left[\partial_t+\bm v_{\bm p} \cdot \nabla_{\bm p}+\bm F(\bm r, \bm p,t)\cdot \nabla_{\bm r}\right] & f(\bm r,\bm p,t)=\\
& = S\{f\}(\bm r,\bm p,t)~,
\end{split}
\end{equation}
where $\bm v_{\bm p}\equiv \nabla_{\bm p} \epsilon_{\bm p}$ is the electron velocity, $\epsilon_{\bm p}$ being the band energy, $\bm F(\bm r, \bm p,t)=-e[\bm E(\bm r,t)+\bm v_{\bm p} \times {\bm B}({\bm r}, t)/c] $ is the total force acting on electrons, ${\bm B}({\bm r},t) = \hat{\bm z}B({\bm r}, t)$ being the external magnetic field, and $S\{f\}(\bm r,\bm p,t)$ is the collision integral. The latter describes all types of electron collisions, 
i.e.~electron-electron, electron-phonon, and electron-impurity collisions.

We solve Eq.~(\ref{eq:Boltzmann}) by using the following Ansatz:
\begin{equation}\label{eq:f1Ansatz}
f(\bm r,\bm p,t)=f_0(\epsilon_{\bm p})-f_0'(\epsilon_{\bm p}) \mathcal{F}(\bm r,\theta_{\bm p},t)~,
\end{equation}
where $f_0(\epsilon)=\{\exp[(\epsilon-\bar{\mu})/(k_{\rm B} T)]+1\}^{-1}$ is the equilibrium Fermi-Dirac distribution function, $f_0'(\epsilon)$ is its derivative with respect to the energy $\epsilon$, and $\theta_{\bm p}$ is the polar angle of the vector $\bm p$. 

Retaining only terms that are linear with respect to $\mathcal{F}(\bm r,\theta_{\bm p},t)$, assuming a uniform and static magnetic field, and Fourier transforming with respect to time, we obtain the following equation for $\mathcal{F}(\bm r,\theta_{\bm p},\omega)$:
\begin{widetext}
\begin{equation}\label{eq:BoltzmannAnsatz}
-i\omega \mathcal{F}(\bm r,\theta_{\bm p},\omega)+\bm v_{\bm p} \cdot \left[\nabla \mathcal{F}(\bm r,\theta_{\bm p},\omega)+e\bm E(\bm r,\omega)\right]
+\omega_{\rm c} \partial_{\theta_{\bm p}} \mathcal{F}(\bm r,\theta_{\bm p},\omega)=S^{\rm el}\lbrace\mathcal{F}\rbrace (\bm r,\theta_{\bm p},\omega) +S^{\rm ee}\lbrace \mathcal{F}\rbrace(\bm r,\theta_{\bm p},\omega)~.
\end{equation}
\end{widetext}
Here $\bm E(\bm r,\omega)$ is the total electric field, i.e.~the sum of the external field and the field generated by the electron distribution itself (the Hartree self-consistent field), $\omega_{\rm c}=eB/(mc)$ is the cyclotron frequency, $m\equiv p_{\rm F}/v_{\rm F}$ is the effective mass, $p_{\rm F}$ and $v_{\rm F}$ being the Fermi momentum and velocity, respectively, $S^{\rm el}\{f\}$ describes momentum non-conserving collision with phonons and impurities, and, finally, $S^{\rm ee}\{f\}$ is the electron-electron collision integral.

We now introduce the Fourier decomposition of the distribution function $\mathcal{F}(\bm r,\theta_{\bm p},\omega)$ with respect to the polar angle:
\begin{equation}\label{eq:Fourier}
\mathcal{F}(\bm r, \theta_{\bm p}, \omega)=\sum_{n=-\infty}^{+\infty}\mathcal{F}_n(\bm r, \omega)e^{in\theta_{\bm p}}~,
\end{equation}
where the Fourier coefficients $\mathcal{F}_n(\bm r, \omega)$ are given by
\begin{equation}
\mathcal{F}_n(\bm r, \omega)=\int \frac{d\theta_{\bm p}}{2\pi}e^{-in \theta_{\bm p}}\mathcal{F}(\bm r, \theta_{\bm p}, \omega)~.
\end{equation}
The lowest-order Fourier coefficients are directly related to simple physical quantities. For example, 
$\mathcal{F}_0$ describes an isotropic dilatation or contraction of the Fermi circle, i.e.~a density perturbation
\begin{equation} \label{eq:Density}
n(\bm r,\omega)\equiv\int d \bm p [f(\bm r,\bm p,\omega)-f_0(\epsilon_{\bm p})]={\cal N}_0 \mathcal{F}_0(\bm r,\omega)~,
\end{equation}
where ${\cal N}_0$ is the density of states at the Fermi energy.

The coefficients $\mathcal{F}_{\pm 1}$ describe a rigid translation of the Fermi surface, which give rise to a finite current:
\begin{equation}\label{eq:Current}
\begin{split}
\bm J(\bm r,\omega) & \equiv\int d \bm p \,\bm v_{\bm p}[f(\bm r,\bm p,\omega)-f_0(\epsilon_{\bm p})]\\
&=\frac{{\cal N}_0v_{\rm F}}{2} 
\begin{pmatrix} \mathcal{F}_{-1}(\bm r,\omega)+\mathcal{F}_1(\bm r,\omega)\\
i\mathcal{F}_1(\bm r,\omega)-i\mathcal{F}_{-1}(\bm r,\omega)\end{pmatrix}~.
\end{split}
\end{equation}

The coefficients $\mathcal{F}_{\pm 2}$ describe an elliptic deformation of the Fermi surface, and are related to the trace-less part of the stress tensor:
\begin{widetext}
\begin{equation}\label{eq:StressTensor}
\begin{split}
{\bm T} (\bm r,\omega) & \equiv\int d \bm p \,\bm p \otimes \bm v_{\bm p}[f(\bm r,\bm p,\omega)-f_0(\epsilon_{\bm p})] =\\
& =\frac{{\cal N}_0v_{\rm F}^2m}{2}
\left[\mathcal{F}_0(\bm r,\omega){\openone}+\frac{\mathcal{F}_2(\bm r,\omega)+ \mathcal{F}_{-2}(\bm r,\omega)}{2}\tau_z+\frac{\mathcal{F}_{-2}(\bm r,\omega)-\mathcal{F}_{2}(\bm r,\omega)}{2i}\tau_x \right]~.
\end{split}
\end{equation}
\end{widetext}
Here, $\openone$ is the $2\times 2$ identity matrix and $\tau_{i}$ with $i=x,y,z$ are ordinary $2\times 2$ Pauli matrices acting on Cartesian indices. Higher-order coefficients describe deformations of the Fermi surface with a more complicated angular dependence.  

Following Ref.~\onlinecite{guo_pnas_2017}, we approximate the collision integrals in Eq.~(\ref{eq:BoltzmannAnsatz}) with the simplest possible expression, which is linear in the coefficients $\mathcal{F}_{n}$ and respects relevant conservation laws.
The collision integral $S^{\rm el}\lbrace \mathcal{F}\rbrace(\bm r,\theta_{\bm p},\omega)$ respects particle number conservation and is described by the phenomenological time scale $\tau$. Its form is
\begin{equation}\label{eq:elColllisionIntegral}
S^{\rm el}\lbrace \mathcal{F}\rbrace(\bm r,\theta_{\bm p},\omega) = -\frac{1}{\tau}\left[\mathcal{F}(\bm r,\theta_{\bm p},\omega)-\mathcal{F}_0(\bm r,\omega)\right]~.
\end{equation}
The electron-electron collision integral should instead respect both particle number and momentum conservation and is described by the parameter $\tau_{\rm ee}$. It reads as following~\cite{guo_pnas_2017,dejong_prb_1995} 
\begin{equation}\label{eq:eeColllisionIntegral}
\begin{split}
& S^{\rm ee} \left\lbrace \mathcal{F} \right\rbrace (\bm r,\theta_{\bm p},\omega)=-\frac{1}{\tau_{\rm ee}}\mathcal{F}(\bm r,\theta_{\bm p},\omega)+\\
& +\frac{1}{\tau_{\rm ee}}  \left[\mathcal{F}_0 (\bm r,\omega)\right.
+ \left.\mathcal{F}_1 (\bm r,\omega)e^{i\theta_{\bm p}}+\mathcal{F}_{-1} (\bm r,\omega)e^{-i\theta_{\bm p}}\right]~,
\end{split}
\end{equation}
where we impose a single relaxation rate due to electron-electron interactions for all non-conserved harmonics, i.e.~${\cal F}_n(\bm r,\omega)$ with $|n|>1$.

Multiplying Eq.~(\ref{eq:BoltzmannAnsatz}) by $e^{-in\theta_{\bm p}}$ and averaging over the angle we get a hierarchy of equations for the moments of the distribution function:
\begin{widetext}
\begin{equation}\label{eq:momentsequation}
\begin{split}
& -i \omega \mathcal{F}_n(\bm r,\omega)+\frac{v_{\rm F}}{2}\left\{\partial_x\left[\mathcal{F}_{n-1}(\bm r,\omega)+\mathcal{F}_{n+1}(\bm r,\omega)\right]-i\partial_y\left[\mathcal{F}_{n-1}(\bm r,\omega)-\mathcal{F}_{n+1}(\bm r,\omega)\right]\right\}\\
& +\frac{ev_{\rm F}}{2}\left[E_x(\bm r,\omega)\left(\delta_{n, 1}+\delta_{n, -1}\right)-iE_y(\bm r,\omega)\left(\delta_{n, 1}-\delta_{n, -1}\right)\right]+in\omega_{\rm c}\mathcal{F}_n(\bm r, \omega)=\\
&-\frac{1}{\tau_{\rm ee}}\left[\mathcal{F}_n(\bm r,\omega)-\mathcal{F}_{0}(\bm r,\omega)\delta_{n, 0} 
-\mathcal{F}_1(\bm r,\omega)\delta_{n, 1}-\mathcal{F}_{-1}(\bm r,\omega)\delta_{n, -1}\right]-\frac{1}{\tau}\left[\mathcal{F}_n(\bm r,\omega)-\mathcal{F}_{0}(\bm r,\omega)\delta_{n, 0}\right]~.
\end{split}
\end{equation}
\end{widetext}
Setting $n=0$ in Eq.~(\ref{eq:momentsequation}) leads to the continuity equation:
\begin{equation}\label{eq:continuitytimedependent}
-i\omega n(\bm r,\omega)+\nabla \cdot \bm J (\bm r,\omega)=0~.
\end{equation}
The two equations for $n=\pm 1$ can be combined to give the Navier-Stokes equation 
\begin{equation}\label{eq:navierstokestimedependent}
\begin{split}
& -i\omega \bm J(\bm r,\omega)  +\frac{1}{m}\nabla \cdot \hat{\bm T}(\bm r,\omega)\\
& +\frac{e v_{\rm F}^2 {\cal N}_0}{2}\bm E(\bm r,\omega)+\omega_{\rm c}\bm J(\bm r,\omega)\times \hat{\bm z}=-\frac{1}{\tau}\bm J(\bm r,\omega)~.
\end{split}
\end{equation}
To obtain a closed set of equations we truncate the series of equations (\ref{eq:momentsequation}) neglecting all the coefficients $\mathcal{F}_n$ with $|n|\geq 3$. By doing this we are able to close the equations for $n=\pm 2$ and we obtain
\begin{equation}
\begin{split}
\mathcal{F}_{\pm 2}(\bm r,\omega)=
-\frac{v_{\rm F}}{2}\frac{(\partial_x \mp i\partial_y )\mathcal{F}_{\pm 1}(\bm r,\omega)}{\frac{1}{\tau}+\frac{1}{\tau_{\rm ee}}-i\omega\pm 2i\omega_{\rm c}}~.
\end{split}
\end{equation}
Replacing this result into the expression for the stress tensor~(\ref{eq:StressTensor}) leads to 
\begin{equation}\label{eq:stresstensor2}
\bm T(\bm r,\omega)=\frac{\mathcal{B}}{\bar{n}}n(\bm r,\omega)\openone-{\bm \sigma}'(\bm r,\omega)~,
\end{equation}
where 
\begin{equation}\label{eq:bulckmodulus}
\mathcal{B}=\frac{\bar{n}mv_{\rm F}^2}{2}=\frac{\bar{n}^2}{\mathcal{N}_0}
\end{equation}
is the bulk modulus~\cite{Giuliani_and_Vignale} of the electron liquid, while the viscous stress tensor is given by 
\begin{equation}\label{eq:viscousstresstensor}
\begin{split}
{\bm \sigma}'(\bm r,\omega)= & m\nu(\omega)
\begin{pmatrix}
\partial_xJ_x-\partial_y J_y & \partial_x J_y+\partial_y J_x\\
\partial_x J_y+\partial_y J_x & -\partial_xJ_x+\partial_y J_y\\
\end{pmatrix}
\\
+ &m \nu_{\rm H}(\omega)
\begin{pmatrix}
\partial_xJ_y+\partial_y J_x & -\partial_x J_x+\partial_y J_y\\
-\partial_x J_x+\partial_y J_y & -\partial_xJ_y-\partial_y J_x\\
\end{pmatrix}.
\end{split}
\end{equation}
This coincides with the expression in Eq.~(\ref{eq:stress}).
Here, the frequency-dependent viscosities are given by
\begin{equation}\label{eq:kinviscosity}
\begin{split}
\nu(\omega)& =\frac{v_{\rm F}^2}{4}\frac{\frac{1}{\tau_{\rm ee}}+\frac{1}{\tau}-i\omega}{\left(\frac{1}{\tau_{\rm ee}}+\frac{1}{\tau}-i\omega\right)^2+4\omega_{\rm c}^2}
\end{split}
\end{equation}
and
\begin{equation}\label{eq:hallviscosity}
\begin{split}
\nu_{\rm H}(\omega) & =-\frac{v_{\rm F}^2}{2}\frac{\omega_{\rm c}}{\left(\frac{1}{\tau_{\rm ee}}+\frac{1}{\tau}-i\omega \right)^2+4\omega_{\rm c}^2}~.
\end{split}
\end{equation}
Setting $\omega=0$ and defining $\nu_0=v_{\rm F}^2\tau_{\rm ee}\tau/[4(\tau_{\rm ee}+\tau)]$, one immediately reaches Eqs.~(\ref{eq:nuB}) and~(\ref{eq:nuHB}) for the magnetic-field-dependent dc viscosities.

\section{Derivation of the BCs reported in Eq.~(\ref{eq:NavierBC})}
\label{app:boundary}

In this Appendix we present a brief derivation of the hydrodynamic BCs in Eq.~(\ref{eq:NavierBC}) for the components of the fluid-element current ${\bm J}$, starting from simple BCs~\cite{reuter_rspa_1948} for the Boltzmann distribution function.

Let us consider a portion of the boundary located at position ${\bm r}_{0}$ and 
a local ``reference system'' defined by the vectors $\hat{\bm e}_{\rm t}$ and $\hat{\bm e}_{\rm n}$ introduced in Sect.~\ref{sec:single}. We denote by $\theta_{0}$ the angle between the tangent vector $\hat{\bm e}_t$ and the $\hat{\bm x}$ direction.
The distribution ${\cal F}(\bm r_0,\theta,\omega)$ represents the density of carriers impinging from the bulk on the boundary if $\theta_0<\theta<\theta_0+\pi$, while it represents the density of carriers scattered from the boundary into the bulk for $\theta_0-\pi<\theta<\theta_0$.
The density of scattered particles is related to the density of impinging particles by
\begin{equation}\label{eq:reflectionprob}
{\cal F}(\bm r_0,\theta+\theta_0,\omega)=\int_0^\pi d\theta' r(\theta,\theta'){\cal F}(\bm r_0,\theta'+\theta_0,\omega)~,
\end{equation}
where $r(\theta, \theta')$ is the probability for a particle impinging with an angle $\theta'$ with respect to the boundary to be scattered with an angle $\theta$, and $-\pi <\theta<0$.
Making use of Eqs.~(\ref{eq:Fourier}) and~(\ref{eq:reflectionprob}), we obtain
\begin{equation}\label{eq:bcFn}
{\cal F}_n(\bm r_0,\omega)e^{in\theta_0}=\sum_{m=-\infty}^\infty e^{im\theta_0}(u_{n-m}+r_{nm}){\cal F}_m(\bm r_0,\omega)~,
\end{equation}
where
\begin{equation}
u_{n}\equiv \int_0^\pi \frac{d\theta}{2\pi}e^{-in\theta}=
\begin{cases}
1/2~\mbox{if}~n=0\\
0~\mbox{if}~n~\mbox{even}\\
-i/(n\pi)~\mbox{if}~n~\mbox{odd}
\end{cases}
\end{equation}
and
\begin{equation}
r_{nm}\equiv\int_{-\pi}^0d\theta e^{-in\theta} \int_0^\pi d\theta' e^{im\theta'}\frac{r(\theta,\theta')}{2\pi}~.
\end{equation}
Since the particle number is conserved in the collisions with the boundary, we have $\int_{-\pi}^0 r(\theta,\theta')d\theta =1$.
This implies
\begin{equation}\label{eq:r0nidentity}
r_{0m}=u_{-m}~.
\end{equation}
Consistently with the procedure followed in deriving the hydrodynamic equations (see Appendix~\ref{app:boltzmann}), we neglect all the contributions stemming from ${\cal F}_n$ with $|n|>2$. Setting $n=0$ in Eq.~(\ref{eq:bcFn}) and making use of Eq.~(\ref{eq:r0nidentity}) yields
\begin{equation}\label{eq:bcn=0}
e^{i\theta_0}{\cal F}_1(\bm r_0,\omega)-e^{-i\theta_0}{\cal F}_{-1}(\bm r_0,\omega)=0~.
\end{equation} 
Noting that 
\begin{equation}
\hat{\bm e}_n \cdot {\bm J}(\bm r_0,\omega)=\frac{{\cal N}_0v_{\rm F}i}{2}\left[e^{i\theta_0}{\cal F}_1(\bm r_0,\omega)-e^{-i\theta_0}{\cal F}_{-1}(\bm r_0,\omega)\right]~,
\end{equation}
we can rewrite Eq.~(\ref{eq:bcn=0}) as
\begin{equation}
\hat{\bm e}_n \cdot {\bm J}(\bm r_0,\omega)=0~.
\end{equation}
Setting $n=1,2$ in Eq.~(\ref{eq:bcFn}) gives instead
\begin{widetext}
\begin{equation}\label{eq:bcn=1}
\left(\frac{i}{\pi}+r_{12}\right)e^{2i\theta_0} {\cal F}_2  +\left(-\frac{1}{2}+r_{11}+r_{1-1}\right)e^{i\theta_0}{\cal F}_1 +\left(-\frac{i}{\pi}  +r_{10}\right){\cal F}_0 +\left(-\frac{i}{3\pi}+r_{1-2}\right)e^{-2i\theta_0}{\cal F}_2 =0
\end{equation}
and
\begin{equation}\label{eq:bcn=2}
\left(-\frac{1}{2}+r_{22}\right)e^{2i\theta_0}{\cal F}_2 +\left(-\frac{4i}{3\pi}+r_{21}+r_{2-1}\right)e^{i\theta_0}{\cal F}_1 +r_{20}{\cal F}_0 +r_{2-2}e^{-2i\theta_0}{\cal F}_2=0~.
\end{equation}
\end{widetext}
In what follows we use a simple one-parameter model for the scattering probability~\cite{reuter_rspa_1948}, which consists in the linear superposition of specular reflection with probability $p$ and diffuse reflection with probability $1-p$.
This reads
\begin{equation}
r(\theta,\theta')=p\delta(\theta+\theta')+\frac{(1-p)}{\pi}~.
\end{equation}
This implies
\begin{equation}\label{eq:rnmcoeff}
r_{mn}=pu_{-n-m}+2(1-p)u_{-n}u_{-m}~.
\end{equation}
Replacing Eq.~(\ref{eq:rnmcoeff}) into Eqs.~(\ref{eq:bcn=1})-(\ref{eq:bcn=2}) and using Eq.~(\ref{eq:bcn=0}) we obtain
\begin{equation}
\begin{pmatrix}
\frac{i(3+p)}{3\pi} & -\frac{i(1+3p)}{3\pi}\\
-\frac{1}{2} & \frac{p}{2}
\end{pmatrix}
\begin{pmatrix}
{\cal F}_2 e^{2i\theta_0}\\
{\cal F}_{-2} e^{-2i\theta_0}
\end{pmatrix}=e^{i\theta_0}{\cal F}_1
\begin{pmatrix}
\frac{1-p}{2}\\
\frac{4i(1-p)}{3\pi}
\end{pmatrix}~.
\end{equation}
Solving for ${\cal F}_{\pm 2}$ gives
\begin{equation}
{\cal F}_2e^{2i\theta_0}=\frac{ie^{i\theta_0}{\cal F}_1(9\pi^2 p -48p-16)}{6\pi (p+1)}
\end{equation}
and
\begin{equation}
{\cal F}_{-2}e^{-2i\theta_0}=\frac{ie^{i\theta_0}{\cal F}_1(9\pi^2 -48-16p)}{6\pi (p+1)}~.
\end{equation}
Using this solution and noting that
\begin{equation}
\begin{split}
&\hat{\bm e}_{\rm t} \cdot [\hat{\bm \sigma}^\prime (\bm r_0,\omega)\cdot \hat{\bm e}_{\rm n} ]=\\
&= -\frac{i{\cal N}_0 v_{\rm F}^2m}{4} \Big[{\cal F}_2(\bm r_0,\omega)e^{2i\theta_0}-{\cal F}_{-2}(\bm r_0,\omega)e^{-2i\theta_0}\Big]
\end{split}
\end{equation}
and
\begin{equation}
\hat{\bm e}_{\rm t} \cdot {\bm J}(\bm r_0,\omega)={\cal N}_0v_{\rm F}e^{i\theta_0}{\cal F}_1(\bm r_0,\omega)~,
\end{equation}
finally leads to Eq.~(\ref{eq:NavierBC}) with
\begin{equation}
\ell_{\rm b}=\frac{6\pi}{9\pi^2-32}\frac{\nu}{v_{\rm F}} \frac{1+p}{1-p}\approx 0.33 \frac{\nu}{v_{\rm F}} \frac{1+p}{1-p}~.
\end{equation}
\section{On the solutions with no-slip BCs}
\label{app:noslip}
In the main text we have focused on the free-surface BCs. 
Here, we discuss how the results for the single-injector setup depend on the BCs, by analysing the no-slip---$\ell_{\rm b}=0$ in Eq. (\ref{eq:NavierBC})---BCs. We start by noting that Eqs.~(\ref{eq:phip}) and~(\ref{eq:phim}), i.e.
\begin{equation}\label{eq:phip_repeated}
\tilde{\phi}_+(k) = I[\tilde{r}_+(k)- i\rho_{\rm H}/k+ 2i \rho_{\nu_{\rm H}} k W^2 +\tilde{r}_{\rm + S}(k)+ \tilde{r}_{\rm + A}(k)]
\end{equation}
and
\begin{equation}\label{eq:phim_repeated}
\tilde{\phi}_-(k) = I[\tilde{r}_-(k)+\tilde{r}_{\rm - S}(k)+ \tilde{r}_{\rm - A}(k)]~,
\end{equation}
hold true {\it independently} of the chosen BCs. In the special case of the no-slip BCs ($\ell_{\rm b}=0$), we find
\begin{widetext}
\begin{align}
\tilde{r}_+(k)&=\rho_0 \frac{q^2\cosh(\bar{k})\sinh(\bar{q})-k q \cosh(\bar{q})\sinh(\bar{k})}{k\{2kq[1-\cosh(\bar{k})\cosh(\bar{q})]+(k^2+q^2)\sinh(\bar{k})\sinh(\bar{q})   \}}~,\\
\tilde{r}_-(k)&=\rho_0 \frac{q[q \sinh(\bar{q})-k \sinh(\bar{k})]}{k\{2kq[1-\cosh(\bar{k})\cosh(\bar{q})]+(k^2+q^2)\sinh(\bar{k})\sinh(\bar{q})\}},\\
\tilde{r}_{\pm {\rm S}}(k)&=0~,\\
\tilde{r}_{+ {\rm A}}(k)&= -i\rho_{\nu_{\rm H}} \frac{q(3\bar{k}^2+\bar{q}^2)[1-\cosh(\bar{k})\cosh(\bar{q})]+k(3\bar{q}^2+\bar{k}^2)\sinh(\bar{k})\sinh(\bar{q})}{2kq[1-\cosh(\bar{k})\cosh(\bar{q})]+(k^2+q^2)\sinh(\bar{k})\sinh(\bar{q})}~,\\
\tilde{r}_{-{\rm A}}(k)&= i\rho_{\nu_{\rm H}}  \frac{ q (\bar{q}^2-\bar{k}^2)  [\cosh(\bar{k}) - \cosh(\bar{q})]}{2kq[1-\cosh(\bar{k})\cosh(\bar{q})]+(k^2+q^2)\sinh(\bar{k})\sinh(\bar{q})}~.\label{eq:CR7}
\end{align}
\end{widetext}
Here, $\bar{k}=kW$, $\bar{q}=qW$, $q=\sqrt{k^2+1/D_\nu^2}$, $D_\nu=\sqrt{\nu \tau}$, 
$\rho_0=m /(\bar{n} e^2 \tau)$, $\rho_{\rm H}=-m \omega_{\rm c}/(\bar{n} e^2 )$, and
$\rho_{\nu_{\rm H}}=m \nu_{\rm H}/(\bar{n} e^2 W^2)$. The resistance $ \tilde{r}_\pm(k)$ coincides with that in the absence of the external magnetic field. Straightforward mathematical manipulations lead to the following asymptotic behavior in the limit $k\to 0$:
\begin{equation}
 \tilde{r}_\pm(k) \to \frac{\rho_0}{k^2 W \{1-2 D_\nu/W\tanh[W/(2D_\nu)]\}}~.
\end{equation}
This means that the asymptotic behavior of the corresponding inverse FT for $|x| \gg W$ is $r_\pm(x) \to -\rho_0 |x|/(2 W)\{1-2 D_\nu/W\tanh[W/(2D_\nu)]\}^{-1}$.

The quantities $\tilde{r}_{\pm {\rm A}}(k)$ are proportional to the kinematic Hall viscosity $\nu_{\rm H}$ and
they are imaginary and odd with respect to the exchange $k\to-k$. 
This implies that the corresponding inverse FTs are odd and real functions of the spatial coordinate $x$.

In the clean $\tau \to \infty$ limit, we find
\begin{align}
 \tilde{r}_+(k) & =-\rho_\nu \frac{2 W \bar{k}[2 \bar{k}+\sinh(\bar{k})]}{1+2 \bar{k}^2-\cosh(2  \bar{k})}~,\\
\tilde{r}_-(k) & =-\rho_\nu \frac{4 W \bar{k}[\bar{k}\cosh(\bar{k})+\sinh(\bar{k})]}{1+2 \bar{k}^2-\cosh(2 \bar{k})}~,\\
\tilde{r}_{+ {\rm A}}(k) & =-i\rho_{\nu_{\rm H}}\frac{4 W  \bar{k}^3}{1+2 \bar{k}^2-\cosh(2  \bar{k})}~,\\
\tilde{r}_{- {\rm A}}(k) & = i\rho_{\nu_{\rm H}}\frac{4 W  \bar{k}^2 \sinh(\bar{k})}{1+2 \bar{k}^2-\cosh(2  \bar{k})}~.\\
\end{align}

Finally, we consider the case of the half-plane geometry, with a current injector placed at the origin~\cite{pellegrino_prb_2016}.
We obtain the solution of the problem for this simple geometry by taking the limit $W\to \infty$ in Eqs.~(\ref{eq:phip_repeated})-(\ref{eq:CR7}). We find
\begin{align}\label{eq:hp}
 \tilde{r}_+(k)  =\rho_0\left[\frac{1}{|k|}+D_\nu^2\left(q+|k|\right)\right]
\end{align}
and
\begin{align}\label{eq:hpA}
\tilde{r}_{+ {\rm A}}(k) & =i \rho_0 \frac{\nu_{\rm H}}{\nu} D_\nu^2\sign(k)\left(q -|k|\right)~,
\end{align}
while $\tilde{r}_-(k) = \tilde{r}_{- {\rm A}}(k) = 0$. Eqs.~(\ref{eq:hp}) and (\ref{eq:hpA}) can be Fourier-transformed analytically. The result is 
\begin{align}
r_+(x) =-\rho_0\left[\frac{1}{\pi}\ln\left(\frac{|x|}{D_\nu}\right)+\frac{D_\nu^2}{\pi x^2} + \frac{D_\nu}{\pi |x|} K_1\left(\frac{|x|}{D_\nu}\right)\right]
\end{align}
and
\begin{align}
r_{+ {\rm A}}(x) & =\rho_0 \frac{\nu_{\rm H}}{\nu} \frac{D_\nu}{2 x}  \left[-I_1\left( \frac{|x|}{D_\nu} \right) +{\bf L}_1\left( \frac{|x|}{D_\nu} \right)\right]~,
\end{align}
where $I_1(x)$ ($K_1(x)$) is the modified Bessel function of first (second) kind and order one and
${\bf L}_{1}(x)$ is the modified Struve function of order one.


\begin{thebibliography}{77}
%
\bibitem{gurzhi_spu_1968}
R.N. Gurzhi, Sov. Phys. Uspekhi~{\bf 11}, 255 (1968).
%
\bibitem{dyakonov_prl_1993}
M. Dyakonov and M. Shur, \href{http://dx.doi.org/10.1103/PhysRevLett.71.2465}{Phys. Rev. Lett.~{\bf 71}, 2465 (1993)}.
%
\bibitem{dyakonov_prb_1995}
M.I. Dyakonov and M.S. Shur, \href{http://dx.doi.org/10.1103/PhysRevB.51.14341}{Phys. Rev. B~{\bf 51}, 14341 (1995)}.
%
\bibitem{dyakonov_ieee_1996}
M. Dyakonov and M. Shur, \href{http://dx.doi.org/10.1109/16.485650}{IEEE Trans. Electron Devices {\bf 43}, 380 (1996)}.
%
\bibitem{conti_prb_1999}
S. Conti and G. Vignale, \href{http://dx.doi.org/10.1103/PhysRevB.60.7966}{Phys. Rev. B~{\bf 60}, 7966 (1999)}.
%
\bibitem{govorov_prl_2004}
A.O. Govorov and J.J. Heremans, \href{http://dx.doi.org/10.1103/PhysRevLett.92.026803}{Phys. Rev. Lett.~{\bf 92}, 026803 (2004)}.
%
\bibitem{muller_prb_2008}
M. M\"{u}ller and S. Sachdev, \href{http://dx.doi.org/10.1103/PhysRevB.78.115419}{Phys. Rev. B~{\bf 78}, 115419 (2008)}.
%
\bibitem{fritz_prb_2008}
L. Fritz, J. Schmalian, M. M\"{u}ller, and S. Sachdev, 
\href{http://dx.doi.org/10.1103/PhysRevB.78.085416}{Phys. Rev. B~{\bf 78}, 085416 (2008)}.
%
\bibitem{muller_prl_2009}
M. M\"{u}ller, J. Schmalian, and L. Fritz, \href{http://dx.doi.org/10.1103/PhysRevLett.103.025301}{Phys. Rev. Lett.~{\bf 103}, 025301 (2009)}.
%
\bibitem{bistritzer_prb_2009}
R. Bistritzer and A.H. MacDonald, \href{http://dx.doi.org/10.1103/PhysRevB.80.085109}{Phys. Rev. B~{\bf 80}, 085109 (2009)}.
%
\bibitem{andreev_prl_2011}
A.V. Andreev, S.A. Kivelson, and B. Spivak, \href{http://dx.doi.org/10.1103/PhysRevLett.106.256804}{Phys. Rev. Lett.~{\bf 106}, 256804 (2011)}.
%
\bibitem{mendoza_prl_2011}
M. Mendoza, H.J. Herrmann, and S. Succi, \href{http://dx.doi.org/10.1103/PhysRevLett.106.156601}{\prl~{\bf 106}, 156601 (2011)}.
%
\bibitem{svintsov_jap_2012}
D. Svintsov, V. Vyurkov, S. Yurchenko, T. Otsuji, and V. Ryzhii, \href{http://dx.doi.org/10.1063/1.4705382}{J. Appl. Phys.~{\bf 111}, 083715 (2012)}.
%
\bibitem{mendoza_scirep_2013}
M. Mendoza, H.J. Herrmann, and S. Succi, \href{http://dx.doi.org/10.1038/srep01052}{Sci. Rep.~{\bf 3}, 1052 (2013)}.
%
\bibitem{tomadin_prb_2013}
A. Tomadin and M. Polini, \href{http://dx.doi.org/10.1103/PhysRevB.88.205426}{\prb~{\bf 88}, 205426 (2013)}.
%
\bibitem{tomadin_prl_2014}
A. Tomadin, G. Vignale, and M. Polini, \href{http://dx.doi.org/10.1103/PhysRevLett.113.235901}{Phys. Rev. Lett.~{\bf 113}, 235901 (2014)}.
%
\bibitem{torre_prb_2015}
I. Torre, A. Tomadin, A.K. Geim, and M. Polini, \href{http://dx.doi.org/10.1103/PhysRevB.92.165433}{Phys. Rev. B~{\bf 92}, 165433 (2015)}.
%
\bibitem{torre_prb_2015_I}
I. Torre, A. Tomadin, R. Krahne, V. Pellegrini, and M. Polini, \href{http://dx.doi.org/10.1103/PhysRevB.91.081402}{Phys. Rev. B~{\bf 91}, 081402(R) (2015)}.
%
\bibitem{narozhny_prb_2015}
B.N. Narozhny, I.V. Gornyi, M. Titov, M. Sch\"{u}tt, and A.D. Mirlin,
\href{http://dx.doi.org/10.1103/PhysRevB.91.035414}{Phys. Rev. B~{\bf 91}, 035414 (2015)}.
%
\bibitem{briskot_prb_2015}
U. Briskot, M. Sch\"{u}tt, I.V. Gornyi, M. Titov, B.N. Narozhny, and A.D. Mirlin,
\href{http://dx.doi.org/10.1103/PhysRevB.92.115426}{Phys. Rev. B~{\bf 92}, 115426 (2015)}.
%
\bibitem{lucas_njp_2015}
A. Lucas, \href{http://dx.doi.org/10.1088/1367-2630/17/11/113007}{New J. Phys.~{\bf 17}, 113007 (2015)}.
%
\bibitem{levitov_naturephys_2016}
L. Levitov and G. Falkovich, \href{http://dx.doi.org/10.1038/nphys3667}{Nature Phys.~{\bf 12}, 672 (2016)}.
%
\bibitem{pellegrino_prb_2016}
F.M.D. Pellegrino, I. Torre, A. K. Geim, and M. Polini
\href{http://dx.doi.org/10.1103/PhysRevB.94.155414}{Phys. Rev. B~{\bf 94}, 155414 (2016)}.
%
\bibitem{lucas_prb_2016}
A. Lucas, J. Crossno, K.C. Fong, P. Kim, and S. Sachdev, 
\href{http://dx.doi.org/10.1103/PhysRevB.93.075426}{Phys. Rev. B~{\bf 93}, 075426 (2016)}.
%
\bibitem{Levchenko_prb_2017}
A. Levchenko, H.Y. Xie, and A.V. Andreev, 
\href{http://dx.doi.org/10.1103/PhysRevB.95.121301}{Phys Rev B.~{\bf 95}, 121301(R) (2017)}.
%
\bibitem{landaufluidmechanics}
L.D. Landau and E.M. Lifshitz, {\it Course of Theoretical Physics: 
Fluid Mechanics} (Pergamon, New York, 1987).
%
\bibitem{avron_prl_1995}
J.E. Avron, R. Seiler, and P.G. Zograf, 
\href{http://dx.doi.org/10.1103/PhysRevLett.75.697}{Phys. Rev. Lett.~{\bf 75}, 697 (1995)}.
%.
\bibitem{tokatly_prb_2007}
I.V. Tokatly and G. Vignale, 
\href{http://dx.doi.org/10.1103/PhysRevB.76.161305}{Phys. Rev. B~{\bf 76}, 161305 (2007)}.
%
\bibitem{tokatly_jpcm_2009}
I.V. Tokatly and G. Vignale, \href{http://dx.doi.org/10.1088/0953-8984/21/27/275603}{J. Phys.: Condens. Matter~{\bf 21}, 275603 (2009)}.
%
\bibitem{read_prb_2009}
N. Read, \href{http://dx.doi.org/10.1103/PhysRevB.79.045308}{Phys. Rev. B~{\bf 79}, 045308 (2009)}.
%
\bibitem{read_prb_2011}
N. Read and E.H. Rezayi, \href{http://dx.doi.org/10.1103/PhysRevB.84.085316}{Phys. Rev. B~{\bf 84}, 085316 (2011)}.
%
\bibitem{haldane_prl_2011}
F.D.M. Haldane, 
\href{http://dx.doi.org/10.1103/PhysRevLett.107.116801}{Phys. Rev. Lett.~{\bf 107}, 116801 (2011)}.
%
\bibitem{hoyos_prl_2012}
C. Hoyos, and D.T. Son,
\href{http://dx.doi.org/10.1103/PhysRevLett.108.066805}{Phys. Rev. Lett.~{\bf 108}, 066805 (2012)}.
%
\bibitem{bradlyn_prb_2012}
B. Bradlyn, M. Goldstein, and N. Read, 
\href{http://dx.doi.org/10.1103/PhysRevB.86.245309}{Phys. Rev. B~{\bf 86}, 245309 (2012)}.
%
\bibitem{sherafati_prb_2016}
M. Sherafati, A. Principi, and G. Vignale, 
\href{http://dx.doi.org/10.1103/PhysRevB.94.125427}{Phys. Rev. B~{\bf 94}, 125427 (2016)}.
%
\bibitem{alekseev_prl_2016}
P.S. Alekseev, \href{http://dx.doi.org/10.1103/PhysRevLett.117.166601}{Phys. Rev. Lett.~{\bf 117}, 166601 (2016)}.
%
\bibitem{cortijo_2DM_2016}
A. Cortijo, Y. Ferreir\'{o}s, K. Landsteiner, and M.A.H. Vozmediano, 
\href{http://dx.doi.org/10.1088/2053-1583/3/1/011002}{2D Mater.~{\bf 3}, 1 (2016)}.
%
\bibitem{scaffidi_prl_2017}
T. Scaffidi, N. Nandi, B. Schmidt, A.P. Mackenzie, and J.E. Moore, 
\href{http://dx.doi.org/10.1103/PhysRevLett.118.226601}{Phys. Rev. Lett.~{\bf 118}, 226601 (2017)}.
%
\bibitem{bandurin_science_2016}
D. Bandurin, I. Torre, R.K. Kumar, M. Ben Shalom, A. Tomadin, A. Principi, G.H. Auton, E. Khestanova, K.S. NovoseIov, I.V. Grigorieva, L.A. Ponomarenko, A.K. Geim, and M. Polini, \href{http://dx.doi.org/10.1126/science.aad0201}{Science~{\bf 351}, 1055 (2016)}.
%
\bibitem{kumar_arxiv_2017}
R.K. Kumar, D.A. Bandurin, F.M.D. Pellegrino, Y. Cao, A. Principi, H. Guo, G.H. Auton, 
M. Ben Shalom, L.A. Ponomarenko, G. Falkovich, I. V. Grigorieva, L.S. Levitov, M. Polini, and A.K. Geim, \href{http://dx.doi.org/10.1038/nphys4240}{Nature Phys. (2017)}.
%
\bibitem{crossno_science_2016}
J. Crossno, J.K. Shi, K. Wang, X. Liu, A. Harzheim, A. Lucas, S. Sachdev, P. Kim, 
T. Taniguchi, K. Watanabe, T.A. Ohki, and K.C. Fong, 
\href{http://dx.doi.org/10.1126/science.aad0343}{Science~{\bf 351}, 1058 (2016)}.
%
\bibitem{moll_science_2016}
P.J.W. Moll, P. Kushwaha, N. Nandi, B. Schmidt, and A.P. Mackenzie, 
\href{http://dx.doi.org/10.1126/science.aac8385}{Science~{\bf 351}, 1061 (2016)}.
%
\bibitem{fugallo_nanolett_2014}
G. Fugallo, A. Cepellotti, L. Paulatto, M. Lazzeri, N. Marzari, and F. Mauri,
\href{http://dx.doi.org/10.1021/nl502059f}{Nano Lett.~{\bf 14}, 6109 (2014)}.
%
\bibitem{capellotti_natcomm_2015}
A. Cepellotti, G. Fugallo, L. Paulatto, M. Lazzeri, F. Mauri, and N. Marzari,
\href{http://dx.doi.org/10.1038/ncomms7400}{Nature Comm.~{\bf 6}, 6400 (2015)}.
%
\bibitem{Giuliani_and_Vignale}
G.F. Giuliani and G. Vignale,
\href{http://dx.doi.org/10.1080/00107510903194710}
{\it Quantum Theory of the Electron Liquid} (Cambridge University Press, Cambridge, 2005).
%
\bibitem{kotov_rmp_2012}
V.N. Kotov, B. Uchoa, V.M. Pereira, F. Guinea, and A.H. Castro Neto, 
\href{http://dx.doi.org/10.1103/RevModPhys.84.1067}{\rmp~{\bf 84}, 1067 (2012)}.
%
\bibitem{steinberg_pr_1958}
M.S. Steinberg, \href{http://dx.doi.org/10.1103/PhysRev.109.1486}{Phys. Rev.~{\bf 109}, 1486 (1958)}. 
%
\bibitem{abanin_science_2011}
D.A. Abanin, S.V. Morozov, L.A. Ponomarenko, R.V. Gorbachev, A.S. Mayorov, M.I. Katsnelson, K. Watanabe, T. Taniguchi, K.S. Novoselov, L.S. Levitov, and A.K. Geim, \href{http://dx.doi.org/10.1126/science.1199595}{Science~{\bf 332}, 328 (2011)}.
%
\bibitem{reuter_rspa_1948}
G.E.H. Reuter and E.H. Sondheimer, \href{http://dx.doi.org/10.1098/rspa.1948.0123}{Proc. R. Soc. Lond. A~{\bf 195}, 336 (1948)}. 
%
\bibitem{principi_prb_2016}
A. Principi, G. Vignale, M. Carrega, and M. Polini,
\href{http://dx.doi.org/10.1103/PhysRevB.93.125410}{Phys. Rev.~{\bf 93}, 125410 (2016)}
%
\bibitem{beconcini_prb_2016}
M. Beconcini, S. Valentini, R.K. Kumar, G.H. Auton, A.K. Geim, L.A. Ponomarenko, M. Polini, and F. Taddei, 
\href{http://dx.doi.org/10.1103/PhysRevB.94.115441}{Phys. Rev. B~{\bf 94}, 115441 (2016)}.
%
\bibitem{taychatanapat_naturephys_2013}
T. Taychatanapat, K. Watanabe, T. Taniguchi, P. Jarillo-Herrero, 
\href{http://dx.doi.org/10.1038/nphys2549}{Nature Phys.~{\bf 9}, 225 (2013)}. 
%
\bibitem{delacretaz_arxiv_2017}
L.V. Delacr\'etaz, and A. Gromov,  \href{https://arxiv.org/abs/1706.03773}{arXiv:1706.03773}.
%
\bibitem{katsnelson} 
M.I. Katsnelson, \href{https://doi.org/10.1017/CBO9781139031080}{\it Graphene: Carbon in Two Dimensions} (Cambridge University Press, Cambridge, 2012).
%
\bibitem{guo_pnas_2017}
H. Guo, E. Ilsevena, G. Falkovich, and L.S. Levitov,
\href{http://dx.doi.org/10.1073/pnas.1612181114}{Proc. Natl. Acad. Sci. (USA)~{\bf 114}, 3068 (2017)}.
%
\bibitem{dejong_prb_1995}
M.J.M. de Jong and L.W. Molenkamp, \href{http://dx.doi.org/10.1103/PhysRevB.51.13389}{\prb~{\bf 51}, 13389 (1995)}.
%
\end{thebibliography}
\end{document}